\documentclass[pra,nofootinbib,twocolumn,floatfix,showpacs,superscriptaddress,groupedaddress]{revtex4}
\usepackage{amsmath,amsfonts,amssymb,bbm,graphicx,subfigure,times}
\usepackage[latin1]{inputenc}

\newcommand{\mathsym}[1]{{}}

\newcommand{\R}{\mathbbm{R}}
\newcommand{\gr}[1]{\boldsymbol{#1}}

\newcommand{\N}{{\cal N}}

\newcommand{\sig}{\gr{\sigma}}
\newcommand{\eq}[1]{Eq.~(\ref{#1})}
\newcommand{\ineq}[1]{Ineq.~(\ref{#1})}
\newcommand{\tr}{{\rm Tr}\,}

\begin{document}

\title{Genuine multipartite entanglement of symmetric Gaussian states: Strong monogamy,
       unitary localization, scaling behavior, and molecular sharing structure}

\author{Gerardo Adesso}
\email{gerardo@sa.infn.it}
\affiliation{Dipartimento di Matematica e Informatica, Universit\`a
degli Studi di Salerno, Via Ponte don Melillo, I-84084 Fisciano (SA), Italy}
\affiliation{CNR-INFM Coherentia, Napoli,
Italy; CNISM, Unit\`a di Salerno; and INFN, Sezione di Napoli -
Gruppo Collegato di Salerno, Italy}

\author{Fabrizio Illuminati}
\email{illuminati@sa.infn.it}
\affiliation{Dipartimento di Matematica e Informatica, Universit\`a
degli Studi di Salerno, Via Ponte don Melillo, I-84084 Fisciano
(SA), Italy} \affiliation{CNR-INFM Coherentia, Napoli, Italy; CNISM,
Unit\`a di Salerno; and INFN, Sezione di Napoli - Gruppo Collegato
di Salerno, Italy} \affiliation{ISI Foundation for Scientific
Interchange, Viale Settimio Severo 65, I-10133 Turin, Italy}

\date{August 18, 2008}

\begin{abstract}
We investigate the structural aspects of genuine multipartite entanglement in Gaussian states of continuous variable systems. Generalizing the results of [Adesso \& Illuminati, Phys. Rev. Lett. {\bf 99}, 150501 (2007)], we analyze whether the entanglement shared by blocks of modes distributes according to a strong monogamy law. This property, once established, allows to quantify the genuine $N$-partite entanglement not encoded into $2,\ldots,K,\ldots,(N-1)$-partite quantum correlations. Strong monogamy is numerically verified, and the explicit expression of the measure of residual genuine multipartite entanglement is analytically derived, by a recursive formula, for a subclass of Gaussian states. These are fully symmetric (permutation-invariant) states that are multi-partitioned into blocks, each consisting of an arbitrarily assigned number of modes. We compute the genuine multipartite entanglement shared by the blocks of modes and investigate its scaling properties with the number and size of the blocks, the total number of modes, the global mixedness of the state, and the squeezed resources needed for state engineering. To achieve the exact computation of the block entanglement we introduce and prove a general result of symplectic analysis: Correlations among $K$ blocks in $N$-mode multi-symmetric and multipartite Gaussian states, that are locally invariant under permutation of modes within each block, can be transformed by a local (with respect to the partition) unitary operation into correlations shared by $K$ single modes, one per block, in effective non-symmetric states where $N-K$ modes are completely uncorrelated. Due to this theorem, the above results, such as the derivation of the explicit expression for the residual multipartite entanglement, its non-negativity, and its scaling properties, extend to the subclass of non-symmetric Gaussian states that are obtained by the unitary localization of the multipartite entanglement of symmetric states. These  findings provide strong numerical evidence that the distributed Gaussian entanglement is strongly monogamous under and possibly beyond specific symmetry constraints, and that the residual continuous-variable tangle is a proper measure of genuine multipartite entanglement for permutation-invariant Gaussian states under any multi-partition of the modes.

\end{abstract}


\maketitle

\section{Introduction}\label{intro}

Although multipartite entanglement is ubiquitous in the physics of many-body quantum systems, it
is very difficult both to characterize and to quantify it \cite{hororev,faziorev}.
While a rather satisfactory understanding has been achieved in the bipartite case \cite{pleniovirmani,vidal}, there is a certain degree of consensus that there is no unique way to define multipartite entanglement, even in the simplest case of pure states \cite{pleniovirmani,eisertgross}. The presence of multipartite entanglement clearly depends on the partitioning that one imposes in order
to group the individual subsystems into parties. Furthermore, given a fixed partition one can single out a hierarchy of different levels of multipartite entanglement, or alternatively $k$-separability, which establishes a smooth connection between the two limiting cases of fully separable state, where the parties are all disentangled, and of fully inseparable states, where entanglement exists across any global bisection, and the parties are supposed to share genuine multipartite entanglement. To complicate things further, not even the concept itself of full inseparability is uniquely defined, as in all quantum systems (starting with the simplest paradigmatic one: a system of three qubits \cite{wstates}) there exist several inequivalent forms and classes of fully inseparable states, with different properties and applications. Therefore, different routes to the definition, characterization, detection, and quantification of multipartite entanglement are possible and have been in fact proposed during the last decade \cite{eisertgross,hororev,pleniovirmani}.

In this context, we would like to recall here three relevant approaches, that are of interest from different perspectives. One possible way, directly generalizing the concept of bipartite entanglement, is to quantify the multipartite entanglement present in a quantum state as the distance between the state itself and the set of fully separable, or (more finely-grained) $k$-separable states, thus adopting an axiomatic approach and a geometric definition of entangled states in terms of their ``distinguishability'' from separable states \cite{weigoldbart,geometrichierarchy}.
Another possibility is to regard entanglement as a physical resource and approach its definition operationally by relating different measures of multipartite and global entanglement to the success or the performance of various multiparty quantum information and communication protocols \cite{masanes,telepoppy,monras}. A further possibility stems from the observation that bipartite entanglement distributes obeying a so-called ``monogamy'' law \cite{terhal}. From the mathematical formulation of such a property, constraints are imposed on genuine multipartite entanglement that lead, in special cases, to suitably defined measures for its quantification \cite{ckw,contangle,yongche,barrylagallina}.

The above, together with other definitions such as quantitative entanglement witnesses \cite{quantentwit} and axiomatic measures like the multipartite squashed entanglement \cite{multisquash}, all come with their burden of mathematical difficulties, e. g. in the solution of the involved optimization problems, and/or with their physical and operational limitations, e.g. in the dimensionality of the systems to which they are applicable, or in the specific protocol they rely on. The various approaches that we have briefly reviewed have nevertheless enabled important progress in the quest for the characterization of multipartite entanglement. Most of these progresses have been achieved for systems of qubits, as the complexity in the description of quantum states scales exponentially with the Hilbert space dimensionality of the constituents. However, the situation becomes more tractable if one moves to algebras of observables with continuous variable (CV) spectra in infinite-dimensional Hilbert spaces \cite{eisplenio,brareview}, and in particular if the analysis is restricted to multimode Gaussian states. In this case, one can derive directly some basic results in analogy with systems of qubits; moreover, some very interesting new features emerge \cite{ourreview}. The obvious advantage, when dealing with Gaussian states (which include states of fundamental physical interest, such as vacuum, coherent, squeezed, and thermal states of harmonic systems) is that they are entirely characterized by the first and second moments of the canonical quadrature phase operators, which delimits the object of the investigation to a finite set of degrees of freedom. Due to this drastic dimensional reduction and to their physical relevance, Gaussian states are still, at present, the theoretical and experimental pillars of quantum information with CVs \cite{covaqial}.

Concerning the multipartite entanglement of Gaussian states, the last two approaches out of the three mentioned above have been particularly successful. From an operational perspective, important basic results were obtained by van Loock and Braunstein \cite{network}. These authors have introduced schemes to produce fully symmetric (i.e. permutation-invariant) $N$-mode Gaussian states exhibiting genuine multipartite entanglement and, accordingly, have devised a communication protocol, the ``quantum teleportation network'',
for the distribution of quantum states exploiting such resources. Subsequently, we have elucidated that the
optimal nonclassical fidelity characterizing the performance of such a protocol (which has been
experimentally demonstrated for $N=3$ \cite{naturusawa}) is in one-to-one correspondence with the presence of genuine multipartite entanglement in fully symmetric Gaussian resources \cite{telepoppy}. This operational characterization of Gaussian entanglement in the framework of quantum teleportation has been followed by
a series of results establishing the (``weak'') monogamy of entanglement for all Gaussian states \cite{contangle,3mpra,hiroshima}. While in its traditional expression the monogamy of entanglement is known to constrain only the bipartite entanglement in different partitions of a multipartite system (as originally proven for qubits \cite{ckw,osborne}), in recent work we have defined a broader framework that postulates the subsistence of a stronger monogamy inequality. This property of ``strong monogamy'' imposes the existence of a trade-off on bipartite {\em and} genuine multipartite entanglement {\em simultaneously} \cite{strong}. We have demonstrated the validity of our construction for the entanglement shared by single modes in the important instance of fully symmetric (pure or mixed) Gaussian states of an arbitrary number $N$ of modes, that constitute the set of entangled resources in the van Loock-Braunstein teleportation network. Remarkably, it is possible to establish that the information-theoretic measure of multipartite entanglement emerging from such a strong monogamy decomposition --- dubbed ``residual contangle'' --- is a monotonic function of the optimal nonclassical fidelity of teleportation for teleportation networks. This result integrates the operational and ``distributional'' approaches to the quantification of genuine multipartite entanglement when applied to multi-party Gaussian states with complete symmetry under permutation of the modes. In passing, we recall that other applications of symmetric, multiparty entangled Gaussian states have been proposed and demonstrated, including secret state sharing \cite{secretsharing} and Byzantine agreement \cite{byz}. More details on these and other applications of multipartite Gaussian states in quantum information can be found in Ref. \cite{brareview,covaqial}, as well as in two recent complementary reviews \cite{ourreview} (theoretical) and \cite{njprev} (experimental).

In the present work we build on the ideas presented in Ref. \cite{strong} and consider the quantification of genuine multipartite entanglement of Gaussian states under less stringent symmetry constraints and more general partitions of the modes, comprising the instance of an arbitrary number of modes per party. We first analyze in general the postulated strong monogamy decomposition:  Specializing to fully symmetric Gaussian states, we derive an analytical formula for the residual contangle. We then show that in such states the strong monogamy decomposition holds for the entanglement shared by arbitrary-sized blocks of modes (in other words, that the corresponding residual contangle is nonnegative), providing strong numerical evidence which extend the domain of validity of the analytical proof of Ref. \cite{strong} (valid for entanglement shared among single modes). This result yields a computable tool that can be readily applied to quantify genuine multipartite entanglement of symmetric Gaussian states. To this aim, we include some handy numerical and analytical-algebraic codes that can be readily computer-implemented. We then investigate in detail how the multipartite entanglement scales with the total number of modes, with the number of parties in which the global system is partitioned, with the size of the individual blocks of modes pertaining to each party, and, finally, with the resource parameters such as the degree of squeezing applied on each mode. This analysis, corroborated by numerous explicit examples, allows to gain a finer understanding of multipartite Gaussian entanglement and to identify its characteristic traits, as well as to single out the optimal conditions in which multipartite entangled Gaussian states can be employed as practical resources for quantum communication tasks.

Here and in the following we focus on Gaussian states of $N$ modes, partitioned into $K$ blocks, dubbed ``molecules'' for ease of visualization \cite{strong}, under the constraint of {\em multi-symmetry}. Imposing the latter property means that we consider states invariant under permutations of modes within each molecule. Clearly, fully symmetric states are those states that are multi-symmetric with respect to any multi-partition. Generalizing a result proven for bipartite bi-symmetric Gaussian entanglement by Serafini, Adesso, and Illuminati \cite{unitarily}, we prove that the ``molecular'' genuine $K$-partite Gaussian entanglement shared by $K$ blocks is {\em unitarily localizable}. Specifically, this means that by performing only local unitary operations independently in the subspace of each single molecule, multi-symmetric $N$-mode Gaussian states are reducible to an equivalent form in which only $K$ modes ($K \leq N$), one per molecule, are correlated --- we call these target modes ``nuclei'' --- while the remaining $(N-K)$ modes become completely uncorrelated. Therefore, the molecular entanglement in the original multi-symmetric $N$-mode Gaussian state across a given $K$-block partition is equal to the nuclear entanglement in the localized Gaussian state of $K$ modes.  Interestingly, such a localized nuclear state does not exhibit, in general, any specific symmetry under permutations of the modes. This fact allows to test explicitly the strong monogamy decomposition \cite{strong} on a restricted class of random pure and mixed non-symmetric multimode Gaussian states as well, obtained by unitarily localizing entanglement in fully symmetric Gaussian states of a higher number of modes with a random structure of partitions. After an extensive performance of such a test, no violation of the strong monogamy constraint has been found. This result provides a first evidence that CV Gaussian entanglement can be strongly monogamous even beyond specific symmetry requirements, and that the integrated operational-distributional approach to the quantification of CV entanglement might possibly lead to a proper measure of genuine multipartite entanglement in general, fully inseparable multimode Gaussian states. On the other hand, more extended tests in more general instances are needed to fully validate such a strong conjecture.

The plan of the paper is as follows. In Sec.~\ref{secprelim} we review some basic notions on Gaussian states of CV systems, some fundamental results on the characterization and quantification of Gaussian entanglement, and some relevant facts on distributed entanglement and monogamy constraints. In Sec.~\ref{secstrong} we briefly discuss the route introduced in Ref.~\cite{strong} to the definition of a measure of genuine multipartite entanglement in CV systems by postulating the existence of a stronger monogamy inequality that constrains both bipartite and multipartite entanglement. In Sec.~\ref{secuniloc} we prove that correlations among blocks of modes in multipartite Gaussian states which are multi-symmetric with respect to local mode permutations can be completely and reversibly localized into effective correlations shared by single modes under a local unitary operation. In Sec.~\ref{secmol} we apply this result and discuss the quantification of genuine multipartite entanglement distributed among molecules of arbitrary size in globally pure or mixed, fully symmetric Gaussian states. We derive an analytical expression for the entanglement and study its scaling properties with the relevant physical parameters, including purity, squeezing, number of modes, number and size of molecules, and the specific multi-partitions imposed on a multimode CV system. In Sec.~\ref{secnuma} we exploit the property of unitary localization to re-interpret our findings in terms of the genuine multipartite entanglement shared by single modes in non-symmetric Gaussian states with a well defined symplectic spectrum. We then perform an extensive numerical investigation which corroborates the measure of genuine multipartite entanglement introduced in Sec.~\ref{secmol} and provides strong evidence that multipartite entanglement is indeed strongly monogamous even in Gaussian states that are not symmetric under permutations of the modes. Conclusions and further outlook are discussed in Sec.~\ref{secconcl}.

\section{Preliminary definitions}\label{secprelim}

\subsection{Gaussian states and Gaussian operations}\label{secPrelim}
Detailed accounts on the mathematical description of Gaussian states in the phase space formalism can be found in many references, e.g. \cite{eisplenio,ourreview}. In view of the subsequent analysis on quantum correlations, here it is sufficient to recall that Gaussian states of a bosonic $N$-mode CV system are completely described in phase space (once the irrelevant first moments are set to zero via local displacements) by the real, symmetric covariance matrix (CM) $\gr{\sigma}$ of the second moments, whose entries are $\sigma_{ij}=1/2\langle\{\hat{X}_i,\hat{X}_j\}\rangle
-\langle\hat{X}_i\rangle\langle\hat{X}_j\rangle$. Here
$\hat{X}=\{\hat x_1,\hat p_1,\ldots,\hat x_N,\hat p_N\}$ is the vector of the field quadrature operators. The canonical commutation relations can be expressed in matrix form: $[\hat
X_{i},\hat X_j]=2i\Omega_{ij}$, with the symplectic form
$\Omega=\oplus_{i=1}^{n}\omega$ and $\omega=\delta_{ij-1}-
\delta_{ij+1},\, i,j=1,2$.  The CM $\gr{\sigma}$ must fulfill the
Robertson-Schr\"odinger uncertainty relation \cite{simon87}
\begin{equation}\label{bonfide}
\gr{\sigma}+i\Omega \geq 0
\end{equation}
in order to describe a physical state. Throughout the paper, $\gr{\sigma}$
will be used indifferently to indicate the CM of a Gaussian state or
the state itself.

Unitary Gaussian operations $U$ amount, in phase space, to
symplectic transformations $S$. The latter are the transformations that preserve the symplectic form,
$\Omega=S^T \Omega S$) acting ``by congruence'' on the CM (i.e. so
that $\sig\mapsto S \sig S^T$). According to Williamson theorem
\cite{williamson36}, the CM of a $N$-mode Gaussian state can be
always brought to diagonal form by means of a global symplectic transformation
(this corresponds to the normal mode decomposition):
\begin{equation}\label{willy}W_{\sig} \sig
W_{\sig}^T = \gr\nu\,,\end{equation}
where $W_{\sig}\in Sp(2N,\R)$ and
$\gr{\nu}=\bigoplus_{k=1}^{N}{\rm diag}\{\nu_k,\,\nu_k\}$ is the CM of a fully uncorrelated state
corresponding to the tensor product of single-mode thermal states.
The quantities $\{\nu_k\}$ are the so-called symplectic eigenvalues
of the CM $\sig$.
A {\em pure} Gaussian state is characterized by $\nu_k=1$,
$\forall\,\,k=1\ldots N$, which implies $\det\sig=1$ and corresponds to the saturation of \ineq{bonfide}. For mixed Gaussian states $\varrho$ with CM $\sig$, the purity $\mu=\tr\varrho^2$ is computable as $\mu=1/\sqrt{\det{\sig}}$.

For what concerns the characterization of bipartite entanglement, the criterion of positivity under partial
transposition (PPT) states that a Gaussian state
$\sig$ is separable (with respect to a $1 \vert N$ bipartition) if
and only if the partially transposed CM $\sig^\Gamma$ satisfies the
uncertainty principle given by \eq{bonfide} \cite{simon,werwolf}.
The suffix ``$\Gamma$'' denotes the partial transposition, implemented by reversing the direction of time in the subspace of either one of the two subsystems in a composite, bipartite CV system \cite{simon}. An ensuing computable measure of CV entanglement is then the \emph{logarithmic negativity}
\cite{vidwer} $E_{\N}\equiv \ln\|{\varrho^\Gamma}\|_{1}$, where $\|
\cdot \|_1$ denotes the trace norm. This measure is an upper bound
to the {\em distillable entanglement} \cite{Bennett96,Rains99} of the state $\varrho$.

We employ a related but different measure of bipartite entanglement: the {\em
contangle} \cite{contangle}, an entanglement monotone under
Gaussian local operations and classical communication (Gaussian LOCC), that
belongs to the family of `Gaussian entanglement measures'
\cite{geof,ordering}. The principal motivation for this choice is that
the contangle, unlike the logarithmic negativity, naturally enables a mathematical treatment of
distributed CV entanglement as emerging from the fundamental
monogamy constraints \cite{contangle,strong}. The contangle
$G_\tau$ is defined for pure states as the square of the logarithmic
negativity and it is extended to mixed states via the Gaussian
convex roof construction \cite{ordering,geof}, that is as the infimum of the
average pure-state entanglement over all decompositions of the mixed
state in ensembles of pure Gaussian states. If $\sig_{i\vert j} $ is
the CM of a (generally mixed) bipartite Gaussian state where
subsystem $i$ comprises one mode only, then the contangle $G_\tau$ can
be computed as \cite{contangle,ourreview}
\begin{equation}
\label{tau} G_\tau (\sig_{i\vert j} )\equiv G_\tau (\sig_{i\vert
j}^{opt} )=g[d_{i\vert j}^2 ],\;\;\;g[x]={\rm arcsinh}^2[\sqrt {x-1}
],
\end{equation}
where $\sig_{i\vert j}^{opt} $ corresponds to a pure Gaussian state,
and $d_{i\vert j} \equiv d(\sig_{i\vert j}^{opt} )=\sqrt {\det
\sig_i^{opt} } =\sqrt {\det \sig_j^{opt} } $, with
$\sig_{i(j)}^{opt} $ the reduced CM of subsystem $i (j)$, obtained
tracing over the degrees of freedom of subsystem $j$ ($i)$. The CM
$\sig_{i\vert j}^{opt} $ denotes the pure bipartite Gaussian state
which minimizes $d(\sig_{i\vert j}^p )$ among all pure-state CMs
$\sig_{i\vert j}^p $ such that $\sig_{i\vert j}^p \le \sig_{i\vert
j} $. If $\sig_{i\vert j} $ is a pure state, then $\sig_{i\vert
j}^{opt} =\sig_{i\vert j} $, while for a mixed Gaussian state
\eq{tau} is mathematically equivalent to constructing the Gaussian
convex roof. For a separable state, $d(\sig_{i\vert j}^{opt} )=1$
which trivially means a vanishing contangle. The contangle $G_\tau$ is completely
equivalent to the Gaussian entanglement of formation \cite{geof},
which quantifies the cost of creating a given mixed, entangled
Gaussian state out of an ensemble of pure, entangled Gaussian
states. Notice also that in general Gaussian entanglement
measures and negativities are inequivalent, in that they may
induce opposite orderings on the set of entangled, mixed non-symmetric
two-mode Gaussian states \cite{ordering}. This result reflects the fact
that different measures of mixed-state entanglement are built on generally
inequivalent operational definitions and thus quantify different `types'
of entanglement.

\subsection{Distributed entanglement and monogamy constraints}

Quantum entanglement, unlike classical correlations,
is said to be {\em monogamous} \cite{terhal}, meaning that it cannot be freely
shared among multiple subsystems of a composite quantum system
(see Ref. \cite{pisa} for a brief topical review).
In its original (we may call it ``weak'') form, the monogamy constraint
imposes the following trade-off on bipartite entanglement distributed among $N$
parties $p_1 \ldots p_N$,
\begin{equation}
\label{ckwine} E^{p_1 \vert (p_2 \ldots p_N
)} \ge \begin{array}{c}\sum_{j \ne 1}^N {E^{p_1 \vert p_j }
}\end{array}\, ,
\end{equation}
where  $E$ is a proper measure of bipartite entanglement
(which may vary depending on the dimensionality of the system under
consideration), and (\ref{ckwine}) is known as the
generalized Coffman-Kundu-Wootters (CKW) inequality \cite{ckw}.
The left-hand side of inequality (\ref{ckwine}) is the bipartite
entanglement between a probe subsystem $p_1$ and the remaining
subsystems taken as a whole. The right-hand side is the total
bipartite block entanglement between $p_1$ and each  of the other
subsystems $p_{j \ne 1}$ in the respective reduced states. The non-negative
difference between these two quantities, minimized over all
choices of the probe subsystem, is referred to as the
\textit{residual multipartite entanglement}: it quantifies the
purely quantum correlations that are not encoded in pairwise form.
In the simplest nontrivial instance of $N=3$, the residual
entanglement has the meaning of the genuine tripartite entanglement
shared by the three subsystems \cite{ckw}. Such a quantity has been
proven to be a tripartite entanglement monotone for pure three-qubit systems \cite{wstates}, when bipartite entanglement is quantified by the tangle \cite{ckw}, and for pure three-mode
Gaussian states, when bipartite entanglement is quantified by the
contangle \cite{contangle}. We recall that the general \ineq{ckwine}
has been established for spin chains ($N$-qubit systems)
\cite{ckw,osborne,yongche} and harmonic lattices ($N$-mode Gaussian states)
\cite{contangle,hiroshima}, with important consequences for the
structure of correlations of those many-body systems
\cite{faziorev,verrucchimany,acinferraro}. A proof of \ineq{ckwine} for arbitrary quantum systems under arbitrary partitions has been also obtained \cite{koashi} when bipartite entanglement is quantified by the squashed entanglement \cite{squash}, a measure which unfortunately lacks computability for almost all quantum states.

\section{Strong monogamy of bipartite and genuine multipartite entanglement}\label{secstrong}

\begin{figure}[bt!]
\includegraphics[width=7cm]{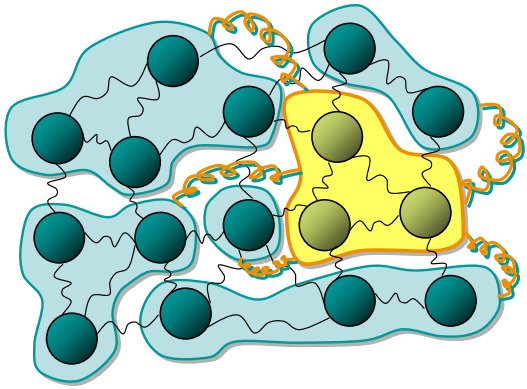} \caption{(Color online)
Pictorial representation of a quantum system multi-partitioned into
$K$ subsystems ($K=6$ in the figure), each containing in general one
or more elementary units (the total number of elementary units, depicted as spheres, is
$N=16$ in the figure). The ``weak'' monogamy of bipartite quantum
correlations imposes that the entanglement between a reference (yellow)
subsystem and the rest of the system, exceeds the sum of the
bipartite entanglements between the reference subsystem and each
other (blue) individual subsystem. In Ref.~\cite{strong} we have
further decomposed this excess between the global block
entanglement and the total bipartite entanglement, into a sum of
genuine multipartite entanglements involving $3$, $4$, ..., $K$
parties at a time. The last contribution quantifies the truly
$K$-partite quantum correlations shared among {\em all} parties. It
is natural to expect that this term is non-negative: such an argument
yields a {\em stronger} monogamy constraint on the simultaneous
distribution of both bipartite and multipartite entanglement.
In Ref.~\cite{strong} we proved the validity of such a
decomposition and hence the existence of such a strong monogamy, for
a system of $N$ harmonic oscillators, one per party, in a globally
permutation-invariant Gaussian state. Here we extend the strong
monogamy framework to multi-partitions comprising an arbitrary number
of modes per party, retrieving a measure of genuine multipartite
entanglement for larger classes of Gaussian states beyond symmetry constraints.}
\label{fiocco}
\end{figure}

In this Section we review the construction introduced in \cite{strong} to refine the right-hand side of \ineq{ckwine} and to recursively extract a quantification of genuine multipartite entanglement via a {\em strong} monogamy constraint on bipartite {\it and} multipartite quantum correlations (see Fig.~\ref{fiocco} for a pictorial representation).

We investigate whether there exists a suitable generalization
of the tripartite CKW-like analysis to an arbitrary number $N$
of modes, such that a {\em genuine} quantifier of $N$-partite entanglement
is naturally derived from a stronger monogamy inequality. We are
motivated by the fact that the residual multipartite entanglement
emerging from the ``weak'' inequality (\ref{ckwine}) includes all
manifestations of $K$-partite entanglement, involving $K$ subsystems
at a time, with $2<K\le N$. Hence, it severely overestimates the
genuine $N$-partite entanglement for $N>3$. It is then compelling to introduce proper
further decompositions of the residual entanglement, by performing
finer investigations on the different ways in which a subsystem can
be entangled with the entire group of the remaining $N-1$ subsystems.

Quite naturally, one can argue that a single subsystem can share individual
pairwise entanglement with each of the other subsystems; and/or genuine
three-partite entanglement involving any two of the remaining subsystems,
and so on, until it can arrive to be genuinely $N$-party entangled with
all the remaining subsystems. We thus introduce the hypothesis that these
contributions are well defined and mutually independent, and verify {\it a posteriori}
that this is indeed the case. Namely, we wish to verify whether in multipartite states,
entanglement is {\it strongly monogamous} in the sense that the
following {\em equality} holds \cite{strong}:
\begin{eqnarray}
\label{strongmono}
E^{p_1|(p_2 \ldots p_N)} &=& \begin{array}{c}\sum_{j=2}^N
E^{p_1|p_j} + \sum_{k>j=2}^N E^{p_1|p_j|p_k} \end{array} \nonumber
\\ &+&  \ldots + E^{\underline{p_1}|p_2|\ldots|p_N}\,,
\end{eqnarray}
where $E^{p_1|p_j}$ is the bipartite entanglement between parties
$1$ and $j$, while all the other terms are multipartite
entanglements involving three or more parties. The last contribution
in \eq{strongmono} is defined implicitly by difference and
represents the residual $N$-partite entanglement. For generally non-symmetric states, one can define a set of residual entanglements $E^{\underline{p_{i_1}}|p_{i_2}|\ldots|p_{i_N}}$ which may possibly differ quantitatively depending on the probe system $p_{i_k}$ with respect to which the total one-to-$N$ bipartite entanglement is decomposed. In order not to overestimate the entanglement shared by all the individual parties, it is thus fairly natural to define the {\em genuine} residual $N$-partite entanglement as the minimum over
all the permutations of the subsystem indexes, \begin{equation}\label{genuine}
E^{p_1|p_2|\ldots|p_N} \equiv
\min_{\{i_1,\ldots,i_N\}}
E^{\underline{p_{i_1}}|p_{i_2}|\ldots|p_{i_N}}\,.\end{equation}
We anticipate that such a definition is also necessary as, for instance, the average or the maximum of the residual entanglements fail to satisfy the strong monogamy condition. For fully symmetric states, there is no need to perform a minimization as the residual entanglements are all equal and independent on the probe subsystems.

\begin{figure*}[t]{
\includegraphics[width=15cm]{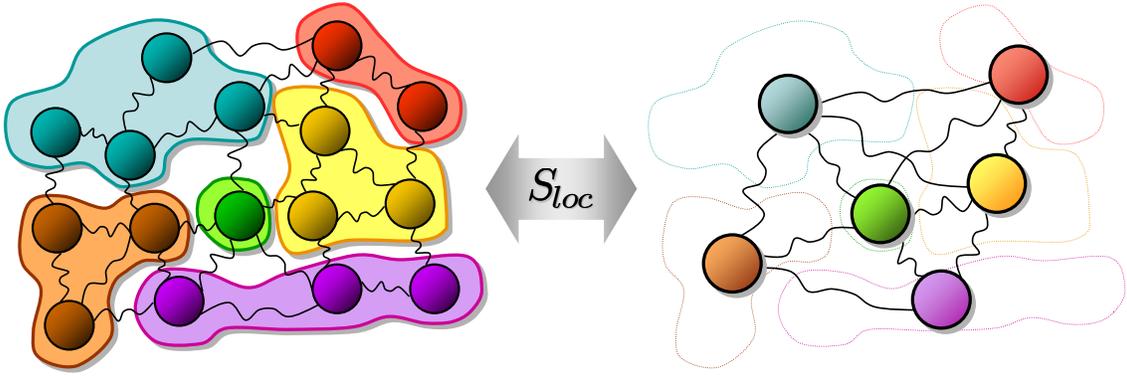} \caption{(Color online)
Pictorial representation of the process of unitary localization and
delocalization of multipartite entanglement in multi-symmetric
Gaussian states. The left picture represents a (pure or mixed)
Gaussian state of $N$ modes ($N=16$ in the figure), all of them in
general correlated, described by a covariance matrix $\sig$. The
$N$-mode system is multi-partitioned into $K<N$ ``molecules'' ${\cal
M}_1,\ldots{\cal M}_K$ ($K=6$ in the figure). Each molecule ${\cal
M}_j$ consists in a block of $m_j$ modes, each of them having the
same reduced covariance matrix (the same color in the figure). By
definition, multi-symmetric states are invariant under mode
permutations within a single molecules. There exists a symplectic
transformation $S_{loc} \equiv W_1 \oplus W_2 \oplus \ldots \oplus
W_K$, acting locally on each molecule, which transforms the original
state into a new one, described by a covariance matrix
$\sig'=S_{loc} \sig S_{loc}^T$, such that all correlations are
concentrated among $K$ modes only (``nuclei''), one per molecule, as
illustrated in the right picture. Precisely, $\sig'$ corresponds to
a generally entangled state of the $K$ nuclei, tensor $(N-K)$
single-mode thermal states which do not bear any correlation with
each other and with the nuclei. Any bipartite and multipartite form
of block entanglement shared by the molecules in the state $\sig$,
is then converted ({\em localized}) into an effective entanglement
among single modes (realizing the transition from left to right in
the figure). The localization is reversible, as $S_{loc}$ amounts to
a local unitary operation at the level of Hilbert space, hence the
entanglement can be redistributed ({\em delocalized}) among all
modes again by applying the inverse symplectic transformation
$S_{loc}^{-1}$ to $\sig'$ (realizing the transition from right to
left in the figure).} \label{figuniloc}}
\end{figure*}

All the contributions to multipartite entanglement appearing in \eq{strongmono}
(except the last one) involve $K$ parties, with $K<N$, and are of
the form $E^{p_1|p_2|\ldots|p_K}$. Each of these terms is defined by
\eq{strongmono} when the left-hand-side is the $1 \vert (K-1)$
bipartite entanglement $E^{p_1|(p_2 \ldots p_K)}$. The $N$-partite
entanglement is thus, at least in principle, {\em computable} in
terms of the known $K$-partite contributions, once \eq{strongmono}
is applied recursively for all $K=2,\ldots,N-1$. To assess
$E^{p_1|p_2|\ldots|p_N}$ as a proper quantifier of $N$-partite
entanglement, one needs first to show its non-negativity on all
quantum states. This property in turn implies that \eq{strongmono}
can be recast as a {\em sharper monogamy inequality}
\begin{equation}\label{strongineq}
E^{p_1|p_2|\ldots|p_N} \ge 0\,,
\end{equation}
constraining both bipartite and genuine $K$-partite ($K \le N$) entanglements in
$N$-partite systems. Such a constraint on distributed entanglement
thus represents a strong generalization of the original CKW inequality
(\ref{ckwine}) \cite{ckw}, implying it, and reducing to it in the special case $N=3$.

In Ref.~\cite{strong} we have further simplified the construction of \eq{strongmono} under the assumption (``full symmetry'') that the states under consideration are invariant under any permutation of the subsystems. Symmetric states play in fact a prominent role in multiparty quantum information applications \cite{definetti} and theoretical entanglement investigations \cite{adescaling,newqubitsnega}. For permutation-invariant Gaussian states we obtained a closed expression for the genuine multipartite entanglement distributed among single modes in terms of an alternating sum of bipartite block entanglements only \cite{strong}. We were moreover able to prove analytically that multipartite entanglement distributed among single-mode parties {\em is} strongly monogamous in arbitrary $N$-mode Gaussian states
that are fully symmetric under arbitrary permutations of the modes \cite{strong}.
We eventually extracted a proper, computable measure of the genuine multipartite entanglement shared by the $N$ single modes in fully symmetric Gaussian states. We proved that the $N$-partite residual entanglement is monotone in the optimal nonclassical fidelity of $N$-party teleportation networks with symmetric Gaussian resources \cite{network,telepoppy}, and thus acquires an operational interpretation and a direct experimental accessibility \cite{strong}.

Here we discuss a more general case in which fully symmetric $N$-mode Gaussian states are partitioned into $K$ blocks of modes --- ``molecules'' --- of arbitrary size. We then analyze the validity of the strong monogamy and the quantification of the genuine multipartite entanglement among the $K$ molecules. A key result on which our analysis is built concerns the possibility of transforming by local (i.e. acting on each molecule) unitary operations, such a $K$-partite block entanglement into the $K$-partite entanglement of an equivalent, generally non-symmetric, Gaussian state of $K$ modes only. We will now show that this property holds indeed and is a consequence of a general result in symplectic analysis.

\section{Unitary localization of multipartite Gaussian entanglement}\label{secuniloc}

Let $\sig^{(N)}_{K:\{m_1,\ldots,m_K\}}$ be the CM of a {\em multi-symmetric}, generally mixed Gaussian state (see Fig.~\ref{figuniloc} for a pictorial representation). With this notation we refer to a Gaussian state of a $N$-mode CV system, multi-partitioned into $K$ subsystems ($K<N$). Each subsystem ${\cal
M}_j$ ($j=1,\ldots,K$) is termed a ``molecule'' and consists in a block of $m_j$ modes. The multi-symmetry (specifically, $K$-symmetry) requirement means that the state is invariant under the permutation of any two modes within each molecule. This implies, in particular, that the $m_j$ modes comprising the $j$-th molecule all have the same marginal covariance matrix. Explicitly,
\begin{widetext}
\begin{eqnarray}\label{multisymcm}
\sig^{(N \equiv \sum_{j=1}^K m_j)}_{K:\{m_1,m_2,\ldots,m_K\}}
&=&\left(
\begin{array}{llll}
 \gr\Sigma_1 & \gr\Gamma_{1,2} & \cdots  & \gr\Gamma_{1,K} \\
 \gr\Gamma^T_{1,2} & \gr\Sigma_2 &  & \gr\Gamma_{2,K}  \\
 \vdots  &  & \ddots & \vdots \\
 \gr\Gamma^T_{1,K} & \gr\Gamma^T_{2,K}  & \cdots & \gr\Sigma_K
\end{array}
 \right)
\\ \nonumber \\ &=&
 \left(
\begin{array}{cccc}
 \overbrace{\left.
\begin{array}{llll}
 \gr\alpha _1 & \gr\varepsilon _1 & \cdots  & \gr\varepsilon _1 \\
 \gr\varepsilon _1 & \ddots & \gr\varepsilon _1 & \vdots  \\
 \vdots  & \gr\varepsilon _1 & \ddots & \gr\varepsilon _1 \\
 \gr\varepsilon _1 & \cdots  & \gr\varepsilon _1 & \gr\alpha _1
\end{array}
\right.}^{m_1} & \left.
\begin{array}{llll}
 \gr\gamma _{1,2} & \cdots  & \cdots  & \gr\gamma _{1,2} \\
 \vdots  & \ddots &  & \vdots  \\
 \vdots  &  & \ddots & \vdots  \\
 \gr\gamma _{1,2} & \cdots  & \cdots  & \gr\gamma _{1,2}
\end{array}
\right. & \left.
\begin{array}{ll}
 \cdots  & \cdots
\end{array}
\right. & \left.
\begin{array}{llll}
 \gr\gamma _{1,K} & \cdots  & \cdots  & \gr\gamma _{1,K} \\
 \vdots  & \ddots &  & \vdots  \\
 \vdots  &  & \ddots & \vdots  \\
 \gr\gamma _{1,K} & \cdots  & \cdots  & \gr\gamma _{1,K}
\end{array}
\right. \\
 \left.
\begin{array}{llll}
 \gr\gamma _{1,2}^T & \cdots  & \cdots  & \gr\gamma _{1,2}^T \\
 \vdots  & \ddots &  & \vdots  \\
 \vdots  &  & \ddots & \vdots  \\
 \gr\gamma _{1,2}^T & \cdots  & \cdots  & \gr\gamma _{1,2}^T
\end{array}
\right. & \underbrace{\left.
\begin{array}{llll}
 \gr\alpha _2 & \gr\varepsilon _2 & \cdots  & \gr\varepsilon _2 \\
 \gr\varepsilon _2 & \ddots & \gr\varepsilon _2 & \vdots  \\
 \vdots  & \gr\varepsilon _2 & \ddots & \gr\varepsilon _2 \\
 \gr\varepsilon _2 & \cdots  & \gr\varepsilon _2 & \gr\alpha _2
\end{array}
\right.}_{m_2} & \left.
\begin{array}{ll}
 \cdots  & \cdots
\end{array}
\right. & \left.
\begin{array}{llll}
 \gr\gamma _{2,K} & \cdots  & \cdots  & \gr\gamma _{2,K} \\
 \vdots  & \ddots &  & \vdots  \\
 \vdots  &  & \ddots & \vdots  \\
 \gr\gamma _{2,K} & \cdots  & \cdots  & \gr\gamma _{2,K}
\end{array}
\right. \\
 \left.
\begin{array}{l}
 \vdots  \\
 \vdots
\end{array}
\right. & \left.
\begin{array}{l}
 \vdots  \\
 \vdots
\end{array}
\right. & \left.
\begin{array}{ll}
 \ddots &  \\
  & \ddots
\end{array}
\right. & \left.
\begin{array}{l}
 \vdots  \\
 \vdots
\end{array}
\right. \\
 \left.
\begin{array}{llll}
 \gr\gamma _{1,K}^T & \cdots  & \cdots  & \gr\gamma _{1,K}^T \\
 \vdots  & \ddots &  & \vdots  \\
 \vdots  &  & \ddots & \vdots  \\
 \gr\gamma _{1,K}^T & \cdots  & \cdots  & \gr\gamma _{1,K}^T
\end{array}
\right. & \left.
\begin{array}{llll}
 \gr\gamma _{2,K}^T & \cdots  & \cdots  & \gr\gamma _{2,K}^T \\
 \vdots  & \ddots &  & \vdots  \\
 \vdots  &  & \ddots & \vdots  \\
 \gr\gamma _{2,K}^T & \cdots  & \cdots  & \gr\gamma _{2,K}^T
\end{array}
\right. & \left.
\begin{array}{ll}
 \cdots  & \cdots
\end{array}
\right. & \underbrace{\left.
\begin{array}{llll}
 \gr\alpha _K & \gr\varepsilon _K & \cdots  & \gr\varepsilon _K \\
 \gr\varepsilon _K & \ddots & \gr\varepsilon _K & \vdots  \\
 \vdots  & \gr\varepsilon _K & \ddots & \gr\varepsilon _K \\
 \gr\varepsilon _K & \cdots  & \gr\varepsilon _K & \gr\alpha _K
\end{array}
\right.}_{m_K}
\end{array}
\right). \nonumber \end{eqnarray}
\end{widetext}

Here $\gr\Sigma_j$ is the reduced (fully symmetric) CM of the block of modes forming the molecule ${\cal M}_j$, while the $\gr\Gamma_{i,j}$ matrices encode the correlations between two molecules ${\cal M}_i$ and ${\cal M}_j$. Each lower-case Greek letter corresponds instead to a two by two matrix.

In Ref.~\cite{unitarily} we have investigated the particular instance of bi-symmetric, bipartite Gaussian states (i.e. the case $K=2$). There, we found that nonlocal correlations in such states are {\em unitarily localizable}: There exist local (i.e. acting independently on each party) and unitary (i.e. symplectic at the CM level) operations that transform the state into a correlated state of two modes only (one per party), tensored with $N-2$ uncorrelated single-mode states. This result stems from the degeneracy of the symplectic spectrum of the reduced (fully symmetric) state of each party \cite{adescaling,unitarily}. Indeed, the proof immediately extends to the more general, multipartite case that we investigate in the present work. The crucial observation is that the local unitary operations enabling the localization are exactly the Williamson transformations $W_j$ which realize the normal mode decomposition, \eq{willy}, in each molecule ${\cal M}_j$. Therefore, they only depend on the molecule on which they act. This entails that localizing operations $W_j$ applied to different molecules commute, and all molecules can be simultaneously diagonalized by a ``local'' (with respect to the given $K$-partition) symplectic operation $S_{loc} \in Sp(2N,\R)$, obtained as the direct sum of the individual Williamson operations $W_j$: $S_{loc} \equiv W_1 \oplus W_2 \oplus \ldots \oplus W_K$. Trivially, each two-molecule reduced state of a multi-symmetric state of the form \eq{multisymcm} is a bi-symmetric state that can be immediately and unitarily localized by the action of $S_{loc}$. Specifically, one then has that the original CM $\sig^{(N)}_{K:\{m_1,\ldots,m_K\}}$, where in principle all modes could have been correlated with each other, is transformed into a new CM
$$\sig'^{(N)}_{K:\{m_1,\ldots,m_K\}}=S_{loc} \sig^{(N)}_{K:\{m_1,\ldots,m_K\}} S_{loc}^T\,.$$ The new CM $\sig'^{(N)}_{K:\{m_1,\ldots,m_K\}}$ corresponds to a Gaussian state in which all correlations are concentrated (localized) among $K$ modes only --- that we term ``nuclei'' --- one per molecule, and the additional $(N-K)$ modes are tensored out.
Explicitly,
\begin{widetext}
\begin{eqnarray}\label{uniloccm}
\sig'^{(N \equiv \sum_{j=1}^K m_j)}_{K:\{m_1,m_2,\ldots,m_K\}}
&=&\overbrace{\left(
\begin{array}{llll}
 \gr\alpha' _1 & \gr\gamma'_{1,2} & \cdots  & \gr\gamma'_{1,K} \\
 \gr\gamma'^T_{1,2} & \gr\alpha'_2 &  & \gr\gamma'_{2,K}  \\
 \vdots  &  & \ddots & \vdots \\
 \gr\gamma'^T_{1,K} & \gr\gamma'^T_{2,K}  & \cdots & \gr\alpha'_K
\end{array}
 \right)}^{\gr{\tilde{\sigma}}'^{(N \equiv \sum_{j=1}^K m_j)}_{K:\{m_1,m_2,\ldots,m_K\}}} \bigoplus_{j=1}^K \gr\nu_j^{\oplus (m_j-1)}
\\ \nonumber \\ &=&
 \left(
\begin{array}{cccc}
 \overbrace{\left.
\begin{array}{llll}
 \gr\alpha' _1 & \gr0 & \cdots  & \gr0 \\
 \gr0 & \gr\nu_1 & \gr0 & \vdots  \\
 \vdots  & \gr0 & \ddots & \gr0 \\
 \gr0 & \cdots  & \gr0 & \gr\nu_1
\end{array}
\right.}^{m_1} & \left.
\begin{array}{llll}
 \gr\gamma'_{1,2} & \gr0 & \cdots  & \gr0 \\
 \gr0 & \gr0 &  & \vdots  \\
 \vdots  & & \ddots & \vdots \\
 \gr0 & \cdots  & \cdots & \gr0
\end{array}\right. & \left.
\begin{array}{ll}
 \cdots  & \cdots
\end{array}
\right. & \left.
\begin{array}{llll}
 \gr\gamma'_{1,K} & \gr0 & \cdots  & \gr0 \\
 \gr0 & \gr0 &  & \vdots  \\
 \vdots  & & \ddots & \vdots \\
 \gr0 & \cdots  & \cdots & \gr0
\end{array}\right. \\
 \left.
\begin{array}{llll}
 \gr\gamma'^T_{1,2} & \gr0 & \cdots  & \gr0 \\
 \gr0 & \gr0 &  & \vdots  \\
 \vdots  & & \ddots & \vdots \\
 \gr0 & \cdots  & \cdots & \gr0
\end{array}\right. & \underbrace{\left.
\begin{array}{llll}
 \gr\alpha' _2 & \gr0 & \cdots  & \gr0 \\
 \gr0 & \gr\nu_2 & \gr0 & \vdots  \\
 \vdots  & \gr0 & \ddots & \gr0 \\
 \gr0 & \cdots  & \gr0 & \gr\nu_2
\end{array}
\right.}_{m_2} & \left.
\begin{array}{ll}
 \cdots  & \cdots
\end{array}
\right. & \left.
\begin{array}{llll}
 \gr\gamma'_{2,K} & \gr0 & \cdots  & \gr0 \\
 \gr0 & \gr0 &  & \vdots  \\
 \vdots  & & \ddots & \vdots \\
 \gr0 & \cdots  & \cdots & \gr0
\end{array}\right. \\
 \left.
\begin{array}{l}
 \vdots  \\
 \vdots
\end{array}
\right. & \left.
\begin{array}{l}
 \vdots  \\
 \vdots
\end{array}
\right. & \left.
\begin{array}{ll}
 \ddots &  \\
  & \ddots
\end{array}
\right. & \left.
\begin{array}{l}
 \vdots  \\
 \vdots
\end{array}
\right. \\
 \left.
\begin{array}{llll}
 \gr\gamma'^T_{1,K} & \gr0 & \cdots  & \gr0 \\
 \gr0 & \gr0 &  & \vdots  \\
 \vdots  & & \ddots & \vdots \\
 \gr0 & \cdots  & \cdots & \gr0
\end{array}\right. & \left.
\begin{array}{llll}
 \gr\gamma'^T_{2,K} & \gr0 & \cdots  & \gr0 \\
 \gr0 & \gr0 &  & \vdots  \\
 \vdots  & & \ddots & \vdots \\
 \gr0 & \cdots  & \cdots & \gr0
\end{array}\right. & \left.
\begin{array}{ll}
 \cdots  & \cdots
\end{array}
\right. & \underbrace{\left.
\begin{array}{llll}
 \gr\alpha' _K & \gr0 & \cdots  & \gr0 \\
 \gr0 & \gr\nu_K & \gr0 & \vdots  \\
 \vdots  & \gr0 & \ddots & \gr0 \\
 \gr0 & \cdots  & \gr0 & \gr\nu_K
\end{array}
\right.}_{m_K}
\end{array}
\right)\, , \nonumber
 \end{eqnarray}
\end{widetext}
where we have denoted by $\gr{\tilde{\sigma}}'$ the effective $K$-mode CM of the nuclei, while the CMs $\gr\nu_j$ represent the uncorrelated single-mode thermal states.
The elements of the CM $\sig'$ can be determined in general from the local (molecular) symplectic invariants of $\sig$ \cite{unitarily}.

>From the previous analysis we see that the property of permutation invariance within each block of modes implies a powerful transformation of multipartite Gaussian states, such that the correlations (including entanglement) shared by the multimode parties can be computed equivalently as correlations among single modes in an effective two-mode Gaussian state. This localization is perfectly reversible, being enabled by the unitary (symplectic) local operation $S_{loc}$, such that the action of $S_{loc}^{-1}$ on $\sig'$ just delocalizes correlations and redistributes them among the various modes of each molecule, reconstructing the original multipartite Gaussian state $\sig$.

An important consequence of the property of unitary localization is that for the relevant subset of Gaussian states that obey the constraint of multi-symmetry, the computation of the multipartite molecular entanglement is greatly simplified by reducing it to the entanglement shared by single modes. For general, non-symmetric multipartite Gaussian states, the minimal form of the CM that can be obtained by local unitary operations involves instead correlations among more than one mode per molecule. Specifically, by virtue of the general decomposition proven in \cite{notsonormal}, there remain irreducible mode-wise and pairwise correlations among the elements of each molecule\footnote{By mode-wise correlations we mean those contained in a state in which single modes each pertaining to a different molecule are correlated. There may be in general a collection of such mode-wise correlated states, involving independently the modes belonging to each molecule. By pair-wise correlations one means instead those correlations contained in a state in which pairs of modes belonging to the same molecule (being themselves correlated) are inter-correlated with pairs of modes belonging to other molecules. More than one pair of modes per molecule can exhibit such correlations in general states.}. It is known that the pairwise correlations vanish for pure, generally non-symmetric Gaussian states \cite{boteroecc}. Here, we have further proved that they vanish as well for generally mixed, multi-symmetric Gaussian states, and that the mode-wise correlations only involve one nuclear mode per party.

We remark that thanks to the unitary localization of Gaussian entanglement, already restricted to the case of bipartite bi-symmetric states \cite{unitarily}, one can immediately conclude that entanglement distillation with correlated copies \cite{jensbrandao} is {\em possible} with Gaussian states and Gaussian operations. This is in sharp contrast to what happens with the standard paradigm of entanglement distillation, where it is known that from several independent and identically distributed copies of mixed bipartite Gaussian states one cannot obtain less copies of more entangled two-mode squeezed states by performing Gaussian LOCCs alone \cite{nogo}. We have just shown instead that the presence of correlations between the copies (testified by a global fully symmetric pure state of the form \eq{multisymcm} with nonzero $\gr\Gamma_{i,j}$'s) enables Gaussian entanglement distillation by simple unitary Gaussian local operations. Finally, it is worth mentioning that results similar to the ones presented in this Section have been demonstrated concerning the possibility of extracting GHZ--type (GHZ stands for Greenberger-Horne-Zeilinger \cite{ghz}) entanglement by local unitary operations performed on multipartite $N$-qubit stabilizer states \cite{fattal}, which are somehow regarded as the discrete variable analogues of Gaussian states \cite{gross}.

\section{Genuine multipartite block entanglement of fully symmetric Gaussian states: molecular hierarchy}  \label{secmol}

Fully symmetric Gaussian states are multi-symmetric with respect to any global multi-partition of the modes. The CM of such states can be written in standard form as \cite{adescaling,unitarily}
\begin{equation}\label{fscm}
\sig_{fs}^{(N)}={\left(%
 \begin{array}{cccc}
  \gr\beta & \gr\zeta & \cdots & \gr\zeta \\
  \gr\zeta & \gr\beta & \gr\zeta & \vdots \\
  \vdots & \gr\zeta & \ddots & \gr\zeta \\
  \gr\zeta & \cdots & \gr\zeta & \gr\beta \\
\end{array}%
\right)}\,,
\end{equation}
with $\gr\beta=\,{\rm diag}\,(b,b)$ ($b \ge 1$) and
$\gr\zeta=\,{\rm diag}\,(z_1,z_2)$. {\em Pure} fully symmetric Gaussian states of $N$ modes are characterized by
$z_{1} = [(b^2-1)(N-2) +\sqrt{(b^{2} -1)(b^2 N^2-(N-2)^2)}]/[2b(N-1)]$, $z_{2} = [(b^2-1)(N-2) +\sqrt{(b^{2} -1)(b^2 N^2-(N-2)^2)}]/[2b(N-1)]$. A {\sl Mathematica} code instruction to construct their CM for any number $N$ of modes is provided by Eq.~(A.2) in the Appendix.

Fully symmetric Gaussian states, \eq{fscm}, are
key resources for essentially all the multiparty CV quantum information protocols implemented so far \cite{brareview}, including the teleportation network \cite{network,naturusawa}, and the Byzantine agreement \cite{byz}. They can be experimentally prepared by sending a single-mode squeezed state with
squeezing $r_m$ in momentum and $(N-1)$ single-mode squeezed states
with squeezing $r_p$ in position, through a network of $N-1$ beam-splitters with tuned transmittivities, as shown in detail in Refs. \cite{network,telepoppy}. Up to local unitaries, the CM of \eq{fscm} can be parameterized equivalently only by considering the average squeezing $r \equiv (r_m+r_p)/2$ needed in the preparation of the state. In fact [code: Eq.~(A.3)], $$b^2=\frac{(N-2) N+2 (N-1) \cosh (4 r)+2}{N^2}.$$
Recall that for $b>1$, i.e. $|r|>0$, the associated fully symmetric Gaussian state is fully inseparable.
In general, the determinant of the reduced $L$-mode CM $\sig_L^{(N)}$ of a fully symmetric $N$-mode pure Gaussian state is given by [code: Eq.~(A.4)]
\begin{equation}\label{detnk}
\det\sig_L^{(N)}=[2 L^2-2 N L+2 (N-L) \cosh (4 r) L+N^2]/N^2.
\end{equation}
Here, the CM $\sig_L^{(N)}$ denotes a fully symmetric Gaussian state, which is mixed as a consequence of the operation of tracing over $N-L$ modes. Recalling that the determinant of the CM $\sig$ is related to the purity $\mu$ of a Gaussian state by $\mu = (\det\sig)^{-1/2}$, we see that the purity of $\sig_L^{(N)}$ strictly decreases with increasing $N$, and is concave as a function of $L$ with a maximum at $L=N/2$ ($N$ even) or $L=(N\pm1)/2$ ($N$ odd).

>From now on, by $\sig^{(N)}_{K:\{m_1,\ldots,m_K\}}$ we will denote a pure, fully symmetric Gaussian state of $N$ modes, multi-partitioned into $K$ molecules, respectively made of $m_1,\ldots,m_k$ modes. Here we allow $\sum_j m_j \equiv M$ to be in principle smaller than $N$, thus considering in general a mixed, symmetric multipartite entangled state of the $K$ molecules. The genuine multipartite entanglement, quantified by the residual contangle as defined via the strong monogamy constraint \eq{strongmono}, has been computed in the case of $m_j=1$ $\forall j$ \cite{strong}, i.e. of mono-modal molecules (``atoms''). We will now show how to evaluate the multipartite entanglement shared by molecules of arbitrary size (multi-modal molecules).

Recursive applications of \eq{strongmono} eventually reduce the task to the computation of all the bipartite $m_i \vert m_j$ entanglements. As we have shown in Sec.~\ref{secuniloc}, the bipartite entanglement between $m_i$ modes and $m_j$ modes in a fully symmetric Gaussian state can be concentrated by local unitary operations onto two modes only (unitary localization) \cite{unitarily}. This effective two-mode Gaussian state turns out to be a mixed state of partial minimum uncertainty, the so-called GLEMS (Gaussian Least Entangled Mixed State) \cite{prl}. A two-mode GLEMS is defined by having the smallest symplectic eigenvalue equal to $1$. This means that one of its normal modes is in the vacuum state (which is trivially a Heisenberg minimum-uncertainty state) and all the mixedness is concentrated in the remaining normal mode. Moreover, a two-mode GLEMS is  completely specified by its global (two-mode) purity and its
two local (single-mode) purities, respectively given by
\begin{eqnarray}\label{abc}
1/c&=&[\det\sig_{m_i+m_j}^{(N)}]^{-1/2}, \nonumber\\ 1/a&=&[\det\sig_{m_i}^{(N)}]^{-1/2}, \\
1/b&=&[\det\sig_{m_j}^{(N)}]^{-1/2}, \nonumber
\end{eqnarray}
where the determinants are defined by \eq{detnk}. The entire family of Gaussian measures of entanglement, including the contangle $G_\tau$ \cite{contangle}, can be computed in closed form for GLEMS \cite{ordering}, yielding [with the notation of \eq{tau}]
\begin{equation}\label{m2glems}d^2({a,b,c})=\left\{
       \begin{array}{ll}
         \frac{\left(a^2-b^2\right)^2}{\left(c^2-1\right)^2}, & c<c_1; \\
         \overset{
            \big[2(a^2+b^2)(c^2+1)+2a^2b^2-a^4-b^4 }{^{-(c^2-1)^2-\sqrt{\delta}\big]/(8c^2)}}
, & c_1\le c<c_2; \\
         1, & c \ge c_2;
       \end{array}
     \right.
\end{equation}
Here $c_1=[(a^2-b^2)^2+2 (a^2+b^2)+\sqrt{(a^2-b^2)^2+8 (a^2+b^2)} |a^2-b^2|]^{1/2}/[2(a^2+b^2)]$, $c_2=\sqrt{a^2+b^2-1}$, and $\delta=(a-b-c-1) (a-b-c+1) (a+b-c-1) (a+b-c+1) (a-b+c-1) (a-b+c+1) (a+b+c-1) (a+b+c+1)$. \eq{m2glems} is coded in the Appendix as Eq.~(A.5).

Therefore, the bipartite contangle between any two molecules ${\cal M}_i$ and ${\cal M}_j$ which are, respectively, $m_i$-modes and $m_j$-modes sub-blocks of a fully symmetric pure $N$-mode Gaussian state, is analytically determined by the expression:
\begin{eqnarray}\label{tauij}
&&\!\!\!\!\!\!\!\!G_\tau (\sig_{m_i\vert m_j}^{(N)})=\\&&\!\!\!\!\!\!\!\!{\rm arcsinh}^2\!\!\!\begin{array}{c}\left[\sqrt {d^2\!\bigg(\sqrt{\det\sig_{m_i}^{(N)}},\!\sqrt{\det\sig_{m_j}^{(N)}},\!\sqrt{\det\sig_{m_i+m_j}^{(N)}}\bigg)\!\!-\!\!1}
\right]\end{array}\!\!,\nonumber
\end{eqnarray}
together with Eqs.~(\ref{detnk}--\ref{m2glems}).

We are now finally in the position to evaluate \eq{strongmono} and determine the genuine $K$-partite entanglement $G_\tau^{res}(\sig_{m_1\vert m_2\vert\ldots\vert m_K}^{(N)})$ distributed among $K$ molecules in a globally pure or mixed fully symmetric Gaussian state of $\sum_{j=1}^K{m_j}\equiv M \le N$ modes. Identifying each subsystem with a molecule, and adopting the contangle as a measure of bipartite entanglement, \eq{strongmono} can be rewritten as
\begin{eqnarray}
\label{monofortemolecolare}
G_\tau^{res}\!\!\!\!\!\!\!&\!\!\!&\!\!\!\!\!(\sig_{m_1\vert m_2\vert\ldots\vert m_K}^{(N)})\\ &=&G_{\tau}(\sig^{(N)}_{m_1|\sum_{j_2>1} m_{j_2}})\nonumber\\ &-& \sum_{j_2>1} G_\tau(\sig^{(N)}_{m_1|m_{j_2}}) \nonumber\\ &-& \sum_{j_3>j_2>1} G_\tau^{res}(\sig_{m_1\vert m_{j_2}\vert m_{j_3}}^{(N)}) \nonumber\\&-& \ldots \nonumber\\ &-& \sum_{j_{K-1}>j_{K-2}>\ldots>j_2>1} G_\tau^{res}(\sig_{m_1\vert m_{j_2}\vert\ldots\vert m_{j_{K-1}}}^{(N)})\nonumber.
\end{eqnarray}
Here we have relabeled the molecules so that their sizes are sorted in ascending order, $m_1 \le m_2 \le \ldots \le m_K$, and we have used the fact that the multipartite entanglement is minimized when the probe subsystem has the smallest determinant among the involved parties \cite{contangle}. \eq{monofortemolecolare} can be easily solved by using a computer language that supports recursion. For this purpose, as well as for other computational tasks of this Section, we have written a {\sl Mathematica} instruction program [Eq.~(A.9) in the Appendix] that evaluates exactly and analytically \eq{monofortemolecolare} for any fixed value of the squeezing $r$, of the number of modes $N$, of the number of partitions $K$, and for each given partition $\{m_1,\ldots,m_K\}$.

The resulting formalism is quite compact, notwithstanding that one is dealing with a complicated combination of nested contributions to multipartite entanglement with alternating signs, and it is by no means apparent
{\it a priori}, even though intuitively expected, that the quantity $G_\tau^{res}(\sig_{m_1 \vert m_2 \vert \ldots \vert m_K}^{(N)})$ is non-negative and thus verifies the postulated property of strong monogamy of multipartite entanglement, \ineq{strongineq}. Although achieving a full analytical proof in the most general situation appears to be a formidable task, we can test the validity of our construction in a wide variety of different situations. In this way, we will single out the peculiar features of multipartite Gaussian entanglement, and its scaling with the number and/or the size of the molecules as well as with the global size of the system and the number of modes that are traced out. As we will show, all investigations confirm the non-negativity of the proposed measure of genuine multipartite entanglement and hence support the validity of the strong monogamy decomposition. In the following, we present a systematic discussion of the various cases and situations considered.

\subsection{Atomic multipartite entanglement} \label{secmolatom}

\begin{figure}[t!]
\includegraphics[width=8.5cm]{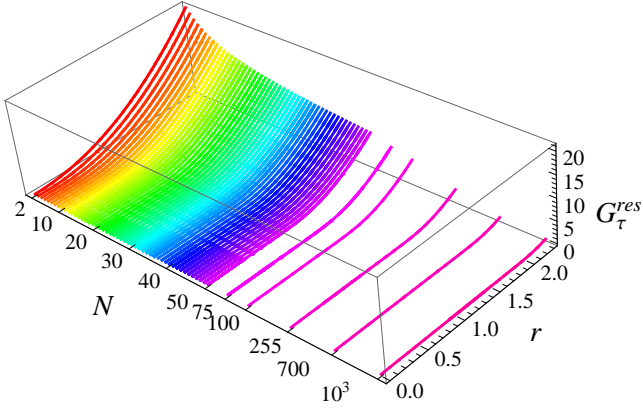} \caption{(Color online)
Genuine multipartite entanglement $G_\tau^{res}$  of pure,
fully symmetric $N$-mode Gaussian states $\sig^{(N)}_{N:\{1,\ldots,1\}}$, plotted as
a function of $N$ (up to $10^3$) and of the average squeezing $r$ needed in the preparation
of the state. All the quantities plotted are dimensionless.}
\label{figatom}
\end{figure}

By the term ``atomic entanglement'' we denote the one shared by individual modes, i.e. by molecules with $m_j=1$ $\forall j=1,\ldots,K$. In this case, one trivially has $\sum_j m_j \equiv M = K$, and the associated $K$-mode state $\sig^{(N)}_{K:\{1,\ldots,1\}}$ is obtained from a pure, fully symmetric Gaussian state of $N$ modes, \eq{fscm}, by tracing over $N-K$ modes. In this specific case, one can derive an explicit formula for $G^{res}_\tau$ and its non-negativity can be proven analytically for any $N$, $K$, and $r$ (the single-mode squeezing) \cite{strong}. We review this important result first, skipping the technical details, for the sake of completeness. The genuine $K$-partite entanglement reads [code: Eq.~(A.7)]
\begin{eqnarray}\label{entj}
&G_\tau^{res}&(\sig^{(N)}_{K:\{1,\ldots,1\}}) \\ &=&\sum_{j=0}^{K-2}\Bigg\{
{\binom{K-1}{j}}(-1)^j \nonumber
\\ \nonumber
&&\times\ {\rm arcsinh}^2
   \left[\frac{2
   \sqrt{K-1-j} \sinh (2
    r)}{\sqrt{N} \sqrt{e^{4
    r} (j+N-K)+K-j}}\right]\Bigg\}\,, \nonumber
\end{eqnarray}
and one can show that $G_\tau^{res}(\sig^{(N)}_{K:\{1,\ldots,1\}}) \ge 0$ (see Fig.~\ref{figatom}), proving the strong monogamy \eq{strongineq} of bipartite and multipartite entanglement distributed among single modes in a fully symmetric Gaussian state \cite{strong}. Ref.~\cite{strong} provides as well an operational interpretation for $G_\tau^{res}$ in the case of pure states ($K=N$) by showing its monotonic dependence on the optimal fidelity of CV teleportation networks \cite{network,telepoppy} for any $N$. Within the scope of this paper, we are interested in studying the scaling of multipartite entanglement with the relevant physical parameters characterizing the multimode Gaussian states.

>From Fig. \ref{figatom} we observe that $G_\tau^{res}$ increases monotonically
with the average squeezing $r$, while it decreases with the number of parties $K$,
eventually vanishing in the field limit $K \rightarrow \infty$. For mixed states
($K$ strictly smaller than $N$), $G_\tau^{res}$ decreases with
the number $N-K$ of the modes that have been traced out, i.e. with the mixedness.
The monotonically decreasing dependence on $K$
can be understood as follows. According to the strong monogamy
decomposition \eq{strongmono}, as the number of modes increases,
the residual non-pairwise entanglement can be encoded in so many different
multipartite forms, that the genuine $K$-partite contribution is
strongly suppressed (frustrated).

\subsection{Molecular multipartite entanglement versus squeezing: promiscuous entanglement distribution}\label{secmolsqz}

>From now on we consider general Gaussian states $\sig^{(N)}_{K:\{m_1,\ldots,m_K\}}$. Recall that $N$ is the total number of modes for a fully symmetric pure Gaussian state $\sig_{fs}^{(N)}$, \eq{fscm}, from which $\sig^{(N)}_{K:\{m_1,\ldots,m_K\}}$ is obtained by tracing over $N-M$ modes (here $M=\sum_j m_j$). The $K$ molecules, i.e. the parties whose shared genuine multipartite entanglement $G_\tau^{res}$ is being computed via \eq{monofortemolecolare}, are formed by $m_1,\ldots,m_K$ modes, respectively.

A unifying trait in our investigations is that $G_\tau^{res}$ is found to be non-negative on any partition
(for a comparison with extended numerical tests, see Sec.~\ref{secnuma}). Moreover, it is a monotonically increasing function of the squeezing $r$ that enters in the engineering of the state $\sig_{fs}^{(N)}$. However, there is a significant difference between the case in which the state of the $K$ molecules is globally pure ($M=N$) and the one in which it is mixed ($M<N$). In the first instance, for any $N$, $K$, and individual molecule sizes $m_j$'s, $G_\tau^{res}$ increases unboundedly with $r$ and diverges as $r \rightarrow \infty$ (which is accompanied by a divergent mean energy). For a mixed molecular state, instead, for any number and size of the molecules the genuine multipartite entanglement remains always finite and saturates at a constant asymptotic value in the limit of infinite squeezing. Examples of these two different behaviors will be illustrated and discussed in the subsequent graphics.

Extensive numerical analysis shows that, as soon as $r>0$, {\em all} possible multi-partitions $\sig^{(N)}_{K:\{m_1,\ldots,m_K\}}$ exhibit genuine multipartite correlations, even though the presence of mixedness limits the maximal amount of achievable entanglement. This coexistence of bipartite, tripartite, \ldots, genuine $K$-partite (etc.) entanglement, together with the fact that all the various contributions are monotonically increasing functions of each other (being all increasing as a function of $r$), defines the so-called ``{\em promiscuous}'' structure of entanglement sharing in Gaussian states \cite{contangle}. Promiscuity is a general feature of distributed entanglement that arises with increasing dimension of the Hilbert space \cite{unlim}. It is fully displayed in Gaussian states of CV systems where despite the firm holding of strong monogamy constraints, there remains a certain freedom in the way quantum information can be shared.

>From an operational point of view, we can conclude this subsection by stating that multi-partitioning the global $N$-mode Gaussian system without discarding any mode is the best choice in order to take full advantage of the squeezing resource and to exploit the infinite entanglement capacity which characterizes CV systems. We will now investigate the dependence of multipartite entanglement on the specific way in which a CV system is multi-partitioned.

\subsection{Molecular multipartite entanglement vs. number of modes}\label{secmolscale}

Here we analyze how the $K$-partite molecular entanglement in the state $\sig^{(N)}_{K:\{m_1,\ldots,m_K\}}$ is affected by a variation of $N$. We have to distinguish between two possibilities: (i) a variation of $N$ with an accompanying variation of the $m_j$'s, i.e. a global re-scaling of the system; (ii) a variation of $N$ keeping fixed molecular sizes, i.e., a variation of the mixedness.

(i) In this case we observe the remarkable phenomenon that the genuine multipartite molecular entanglement of symmetric Gaussian states is {\em scale invariant}. Let us consider the state $\sig^{(N)}_{K:\{m_1,\ldots,m_K\}}$, and construct another state $\sig^{(sN)}_{K:\{s\,m_1,\ldots,s\,m_K\}}$ where both the total number of modes $N$ and the sizes of the individual molecules $m_1,\ldots,m_K$ (and hence their sum $M$) are re-scaled by an arbitrary integer factor $s$. It is immediate to see that the genuine multipartite entanglement $G^{res}_\tau$ among the $K$ molecules is equal in both states. This follows by recalling from Eqs.~(\ref{abc}--\ref{tauij}) that $G^{res}_\tau$ depends ultimately only on the reduced determinants $\det\sig_L^{(N)}$ of $L$-mode sub-blocks of the $N$-mode state, and by observing that $\det\sig_{s L}^{(s N)}$ in \eq{detnk} is independent of the scale factor $s$.  Therefore we see that {\em size does not matter} for what concerns multipartite Gaussian entanglement, if the partition structure is preserved. This is relevant in view of the practical exploitation of
Gaussian states for communication tasks \cite{brareview}: Adding redundance, e.g. by doubling the size of the individual subsystems, yields no advantage for the multiparty-entangled resource.

It is interesting to observe that an analogous property of scale invariance holds as well for the set of hierarchical, multi-component measures of geometric entanglement in arbitrary, fully symmetric pure states of $N$-qubit systems \cite{geometrichierarchy}. The fact that two different measures of genuine multipartite entanglement evaluated on two very different systems ($N$-qubit systems and $N$-mode CV systems) exhibit the same structure of scale invariance hints at the intriguing possibility that multipartite entanglement {\it is} indeed scale invariant on all fully symmetric states of arbitrary multi-party quantum systems.

\begin{figure}[t!]
\includegraphics[width=8.5cm]{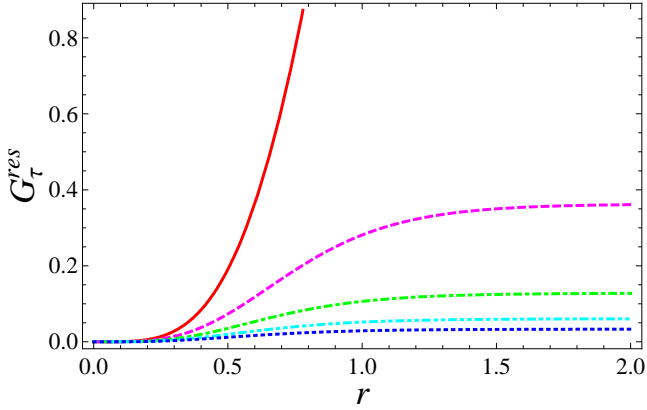}
\caption{(Color online) Genuine multipartite entanglement $G^{res}_\tau$
in the state $\sig^{(N)}_{4:\{1,2,3,4\}}$ as a function of the squeezing $r$.
The plot reports the entanglement shared by four molecules made respectively of $1$, $2$, $3$, and $4$ modes in $10$-mode fully symmetric Gaussian states, obtained, respectively, as reductions of a pure, fully symmetric $N$-mode Gaussian state with $N=10$ (solid red line, no modes are traced out),
$N=11$ (dashed magenta line),  $N=12$ (dot-dashed green line), $N=13$ (dot-dot-dashed cyan
line), and $N=14$ (dotted blue line). The genuine multipartite molecular entanglement decreases with increasing $N$, i.e. with the number of discarded modes. In the pure-state instance (no modes being discarded) it diverges as $r \rightarrow \infty$. All the quantities plotted are dimensionless.} \label{fiversusn}
\end{figure}

(ii) On the other hand, if the number of modes $N$ is increased to $N'>N$ without re-scaling the size of the molecules, this results in a state $\sig^{(N')}_{K:\{m_1,\ldots,m_K\}}$ which is always less pure than the state $\sig^{(N)}_{K:\{m_1,\ldots,m_K\}}$. This is because we are increasing the number of modes which are traced out. As it could be intuitively expected, in this case the genuine $K$-partite molecular entanglement decreases monotonically with $N'$. An example of this behavior when scaling the number of modes is illustrated in Fig.~\ref{fiversusn}.

\subsection{Molecular multipartite entanglement vs. number of molecules}\label{secmolk}

Next, we keep both $N$ and $M$ fixed, and study the dependence of $G_\tau^{res}$ on the number $K$ of molecules in which the $M$ modes in the state $\sig^{(N)}_{K:\{m_1,\ldots,m_K\}}$ are partitioned. To visualize the problem, suppose to have pure states ($M=N$) with $M$ large enough to allow the system to be partitioned into $K_1$ or $K_2$ $\ldots$ or $K_M$ equally sized molecules\footnote{Although our findings do not rely on these assumptions, we consider pure states and molecules with the same size for simplicity in order to enable a more direct comparison, since the multipartite entanglement is further affected by the degree of mixedness (see previous subsections) and by the size of the individual molecules (see next subsection).} as well as into $M$ atoms (each molecule formed by one mode). The integer $M$ can be chosen as a common multiple of the set $\{K_1,\ldots,K_M\}$, in such a way that in each instance the system is partitioned into $K_i$ molecules made of $M/K_i$ modes. This is an interesting case to study as we may imagine the experimental situation in which a given multimode symmetric Gaussian state has been prepared, and an option is available whether to split and distribute it to few or to many parties, in order for them to share as much entanglement as possible.

\begin{figure}[t!]
\includegraphics[width=8.5cm]{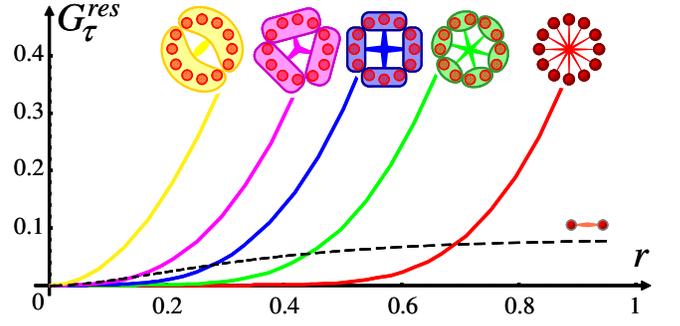}
\caption{(Color online) Molecular entanglement hierarchy in a
pure fully symmetric $M$-mode Gaussian state with $M=12$. From left to right: the
bipartite entanglement between two $6$-mode molecules (yellow line)
is strictly stronger than the genuine tripartite entanglement
between three $4$-mode molecules (magenta line), and so on, while
the atomic genuine $12$-partite entanglement (red line) is the
weakest one. The bipartite entanglement between any two modes in the reduced mixed state obtained by tracing over $10$ modes is plotted as well as reference
(dashed black line). All the quantities plotted are dimensionless.} \label{figcul}
\end{figure}

By virtue of the scale invariance of $G_\tau^{res}$ (see Sec.~\ref{secmolscale}), the genuine multipartite entanglement shared by $K_i$ even-sized molecules in the state $\sig^{(M)}_{K_i:\{M/K_i,\ldots,M/K_i\}}$ is equal to the genuine multipartite entanglement shared by $K_i$ modes in the state $\sig^{(K_i)}_{K_i:\{1,\ldots,1\}}$, i.e. to the atomic entanglement of the latter state. We have shown in Sec.~\ref{secmolatom} that the atomic entanglement, \eq{entj}, monotonically decreases with an increasing number of entangled modes, $K_i$ in this case. These two observations allow to conclude that the genuine multipartite molecular entanglement in the state $\sig^{(M)}_{K_i:\{M/K_i,\ldots,M/K_i\}}$ is monotonically decreasing with an increasing number $K_i$ of molecules. As a general rule, therefore, a smaller number of larger molecules shares strictly more entanglement than a larger number of smaller molecules: The maximum entanglement is always obtained by bisecting symmetrically the system and distributing it to two parties only. An example of this hierarchy of molecular entanglements is illustrated and discussed in Fig.~\ref{figcul}.

\subsection{Multipartite entanglement vs. size of the molecules}\label{secmolmpart}

Finally, we keep $N$, $M$, and $K$ fixed. We are thus dealing with a well defined fully symmetric Gaussian state of $M$ modes, with an assigned degree of mixedness determined by the number $N-M$ of discarded modes, and, importantly, with a well defined number of molecules $K$ in which the $M$ modes are partitioned. This instance corresponds to the experimental situation in which a protocol involving $K$ parties has to be implemented, and  $\sig^{(N)}_{K:\{m_1,\ldots,m_K\}}$ is the resource state that needs to be employed. There is a freedom left in the choice of the individual sizes $m_j$ of each molecule (constrained to $\sum_j m_j=M$), which corresponds to the freedom in assigning a certain number of modes to each party. The distribution can easily break the symmetry of the original state whenever the different parties are assigned a different number of modes $m_j$. We thus wish to investigate how the genuine multipartite entanglement shared by the $K$ molecules, \eq{monofortemolecolare}, depends on the partition $\{m_1,\ldots,m_K\}$, i.e. on the specific molecular size.

Determining all the possible independent partitions of $M$ modes into $K$ subsets is a combinatorial problem. Here only the individual values of the integers $m_j$ are relevant, not their order (e.g. $\{1,2,3\}$ and $\{3,2,1\}$ count as the same partition).
It is thus convenient to introduce a ranking of the different partitions in order to be able to write a sorted list, without repetitions, of all the possible sets $\{m_1,\ldots,m_K\}$ given any integer value of $M$ and $K$. We denote by ${\cal P}_{M,K}=\{p_{M,K}^{[i]}\}_{i=1}^{\aleph_{M,K}}$ the list of partitions, whose $i$-th element is of the form $p_{M,K}^{[i]}=\{m_1^{[i]},\ldots,m_K^{[i]}\}$; the cardinality of ${\cal P}_{M,K}$ is the integer $\aleph_{M,K}$.
The ranking we choose is very natural. Let $\eta_{M,K}^{[i]}$ be the integer decimal number whose digits are given by the $m_j^{[i]}$'s in base $M$, i.e. $\eta_{M,K}^{[i]} = \sum_{j=1}^K m_j^{[i]} M^{K-j}$. Then the elements $p_{M,K}^{[i]}$ of the list ${\cal P}_{M,K}$ are sorted in such a way that the corresponding ranking numbers $\eta_{M,K}^{[i]}$ are in ascending order. A simple way to construct the sorted list of partitions of $M$ modes into $K$ molecules is given by our {\sl Mathematica} instruction {\it KSortedPartitions[$M$, $K$]}, coded as
Eq.~(A.8) in the Appendix. For instance, ${\cal P}_{8,4}=\big\{\{1, 1, 1, 5\}, \{1, 1, 2, 4\}, \{1, 1, 3, 3\}, \{1, 2, 2, 3\}, \{2, 2, 2, 2\}\big\}$.

\begin{figure*}[tb!]{\centering{
\subfigure[] {\includegraphics[width=7.8cm]{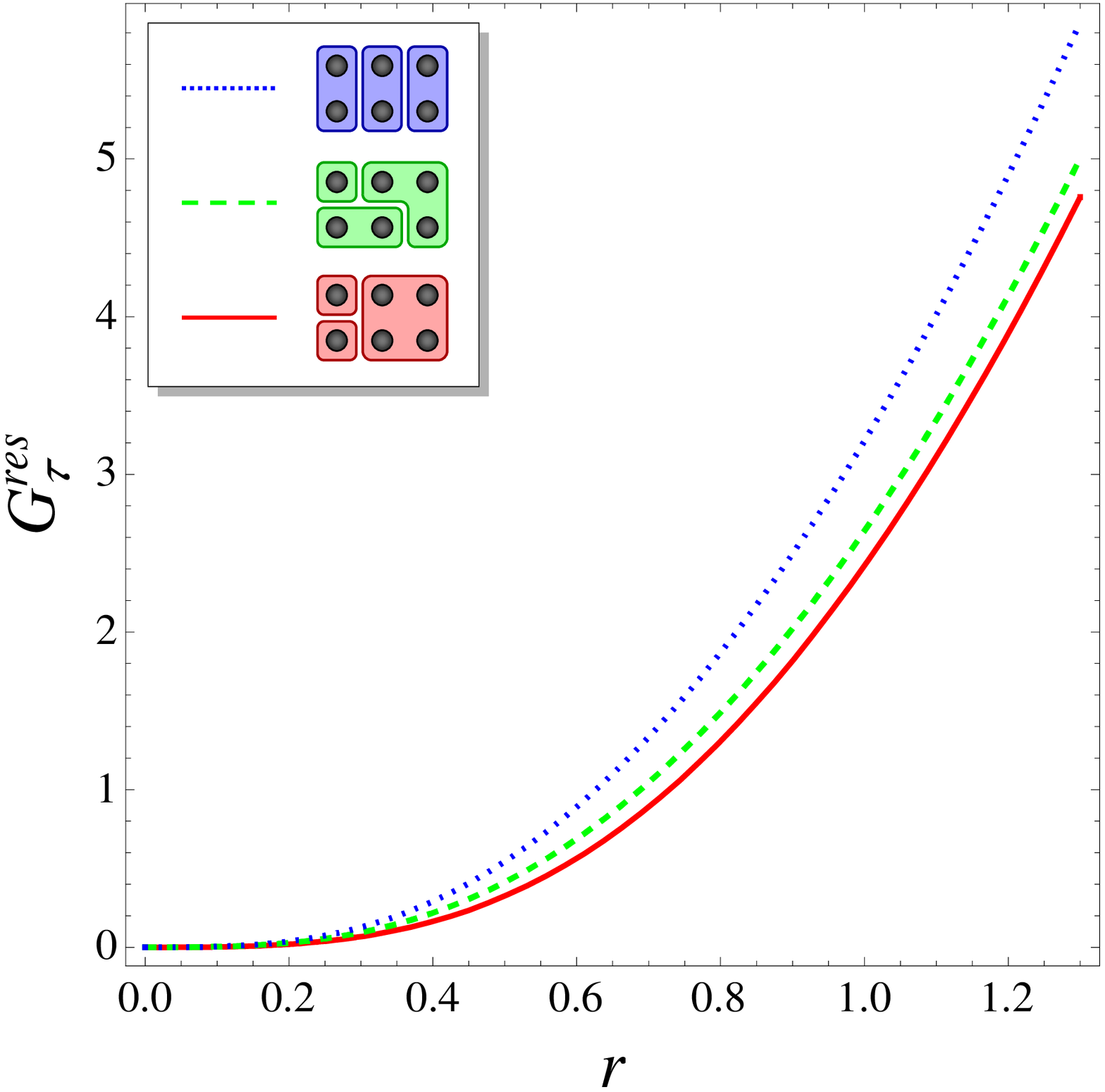}}
\hspace*{0.5cm} \subfigure[]
{\includegraphics[width=7.8cm]{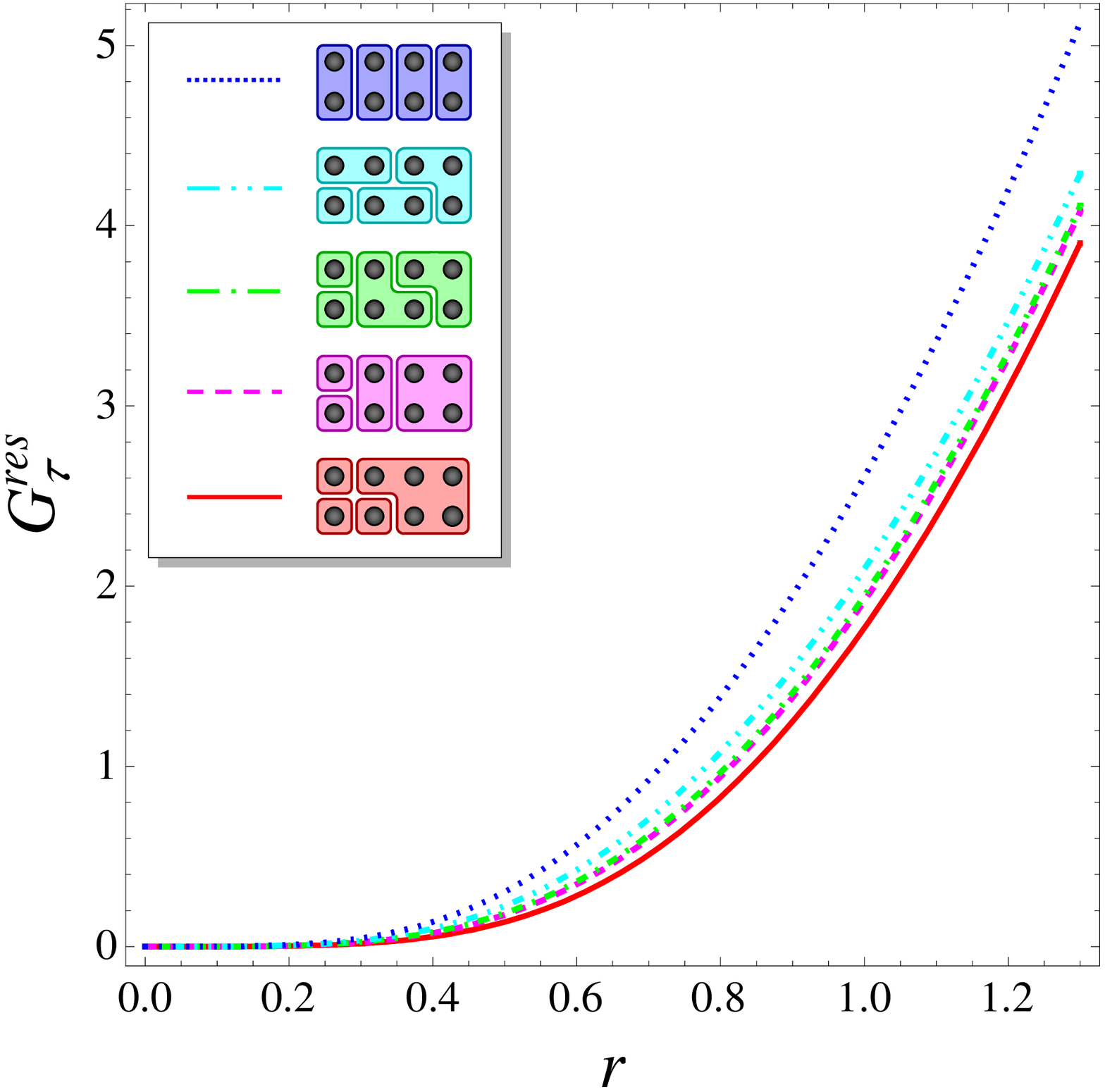}} \\
\vspace*{0.4cm} \subfigure[]
{\includegraphics[width=7.8cm]{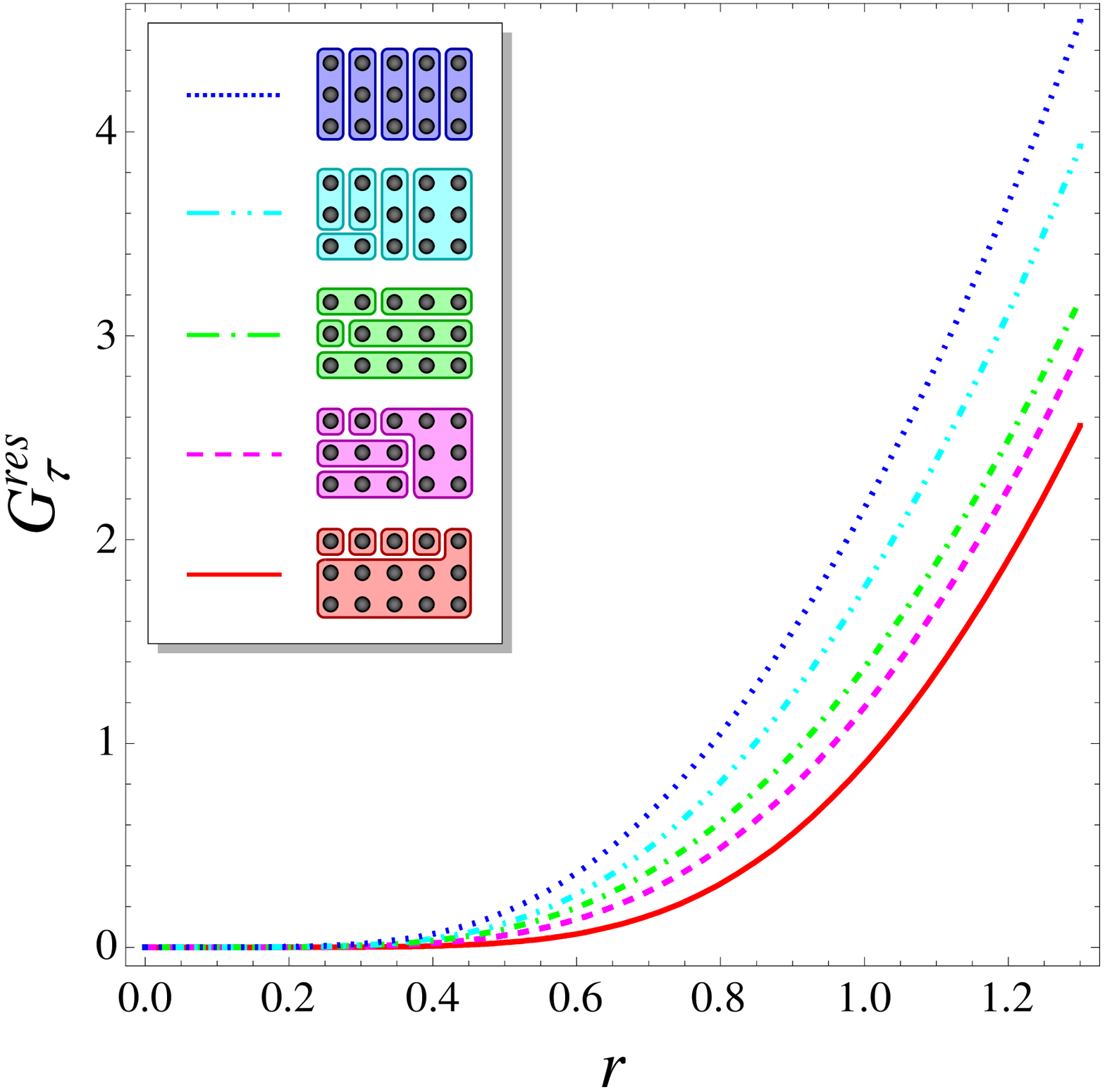}} \hspace*{0.5cm}
\subfigure[] {\includegraphics[width=7.8cm]{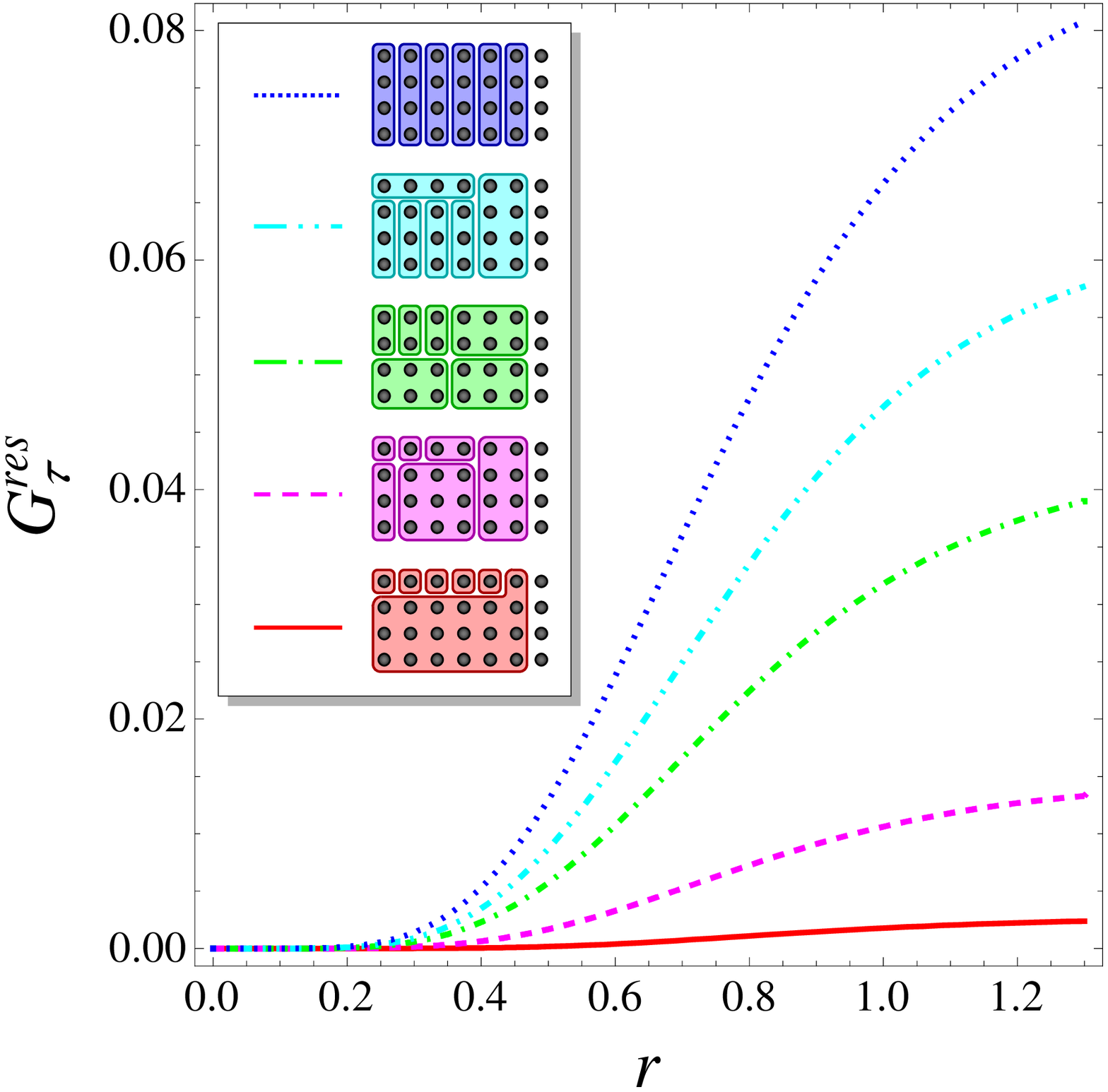}}
\caption{(Color online) Plot, as a function of the squeezing $r$, of
the genuine $K$-partite entanglement $G^{res}_\tau$ in $N$-mode pure
fully symmetric Gaussian states, multipartitioned into $K$ molecules
of varying sizes $\{m_1,\ldots,m_K\}$, with $\sum_j m_j=M$ (see plot legends for a pictorial representation
of the molecular groupings). Panel (a) depicts the instance $M=N=6$,
$K=3$; the molecules are made of, from bottom to top, $\{1,1,4\}$
modes (solid red line), $\{1,2,3\}$ modes (dashed green line), and
$\{2,2,2\}$ modes (dotted blue line), respectively. Panel (b)
illustrates the instance $M=N=8$, $K=4$; the molecules are made of, from
bottom to top, $\{1,1,1,5\}$ modes (solid red line), $\{1,1,2,4\}$
modes (dashed magenta line), $\{1,1,3,3\}$ modes (dot-dashed green
line), $\{1,2,2,3\}$ modes (dot-dot-dashed cyan line), and
$\{2,2,2,2\}$ modes (dotted blue line), respectively. Panel (c)
depicts the instance $M=N=15$, $K=5$; the molecules are made of, from
bottom to top, $\{1,1,1,1,11\}$ modes (solid red line),
$\{1,1,1,3,7\}$ modes (dashed magenta line), $\{1,2,3,4,5\}$ modes
(dot-dashed green line), $\{2,2,2,3,6\}$ modes (dot-dot-dashed cyan
line), and $\{3,3,3,3,3\}$ modes (dotted blue line), respectively
(only 5 out of 30 possible molecular partitions are shown). Panel
(d) illustrates a mixed-state instance: the total number of modes in the
pure state is $N=28$ but only $M=24$ of them are partitioned into
$K=6$ molecules (i.e. we are tracing over the extra $4$ modes to be
left with the multipartite molecular entanglement in a $24$-mode
fully symmetric mixed state); the molecules are made of, from bottom
to top, $\{1,1,1,1,1,19\}$ modes (solid red line), $\{1,1,2,3,8,9\}$
modes (dashed magenta line), $\{2,2,2,6,6,6\}$ modes (dot-dashed
green line), $\{3,3,3,3,4,8\}$ modes (dot-dot-dashed cyan line), and
$\{4,4,4,4,4,4\}$ modes (dotted blue line), respectively (only 5 out
of 199 possible molecular partitions are shown). Notice how, for a
fixed total number of modes and parties (molecules), the distributed
multipartite entanglement increases with the ranking of the sorted list
of partitions, i.e. it is bigger whenever the individual molecule
sizes are more similar to each other. Notice also how the genuine
multipartite entanglement in any partition increases unboundedly
with the squeezing $r \gg 0$ for pure states [Panels (a)--(c)],
while it eventually saturates for mixed states [Panel (d)]. All the quantities plotted are dimensionless.}
\label{figreg}}}
\end{figure*}

Having determined the combinatorial structure of the possible independent partitions, we can investigate the dependence of the genuine multipartite entanglement on the molecular partitioning for a rich variety of pure and mixed fully symmetric Gaussian states with different assigned values of $N$, $M$, and $K$. Some representative instances are illustrated in Fig.~\ref{figreg}. After an extensive numerical analysis, we have found, quite interestingly, that $G_\tau^{res}$ appears to increase monotonically with increasing ranking of the sorted list of all the possible partitions. Specifically, the most inefficient way to partition the system is always the most unbalanced one that assigns a single mode to all parties but one, and all the remaining modes to the last party. On the contrary, the best strategy to distribute the maximum $K$-partite entanglement corresponds to a balanced partitioning, in which all the parties have almost the same number of modes (i.e. all equal $m_j$'s or the closest possible setting, depending on $M$ being even or odd and on its ratio with $K$). Between these two extremal cases lie all the intermediate situations in which, gradually, the molecules tend to assume similar sizes (following the ranking of the list ${\cal P}_{M,K}$), and the genuine $K$-partite entanglement increases accordingly.

The overall picture shows that $G_\tau^{res}$ exhibits a hierarchical structure even with respect to the way the system is partitioned. In fact, such a picture matches the intuition that molecules with similar sizes yield a stronger entanglement. It is quite natural indeed to expect that if the different parties share high genuine multipartite entanglement, they should carry rather similar amounts of quantum information (quantifiable by the local purity), which is precisely the case for molecular partitions with high ranking in our sorted list.

\section{Numerical test of strong monogamy. Genuine multipartite entanglement of effective non-symmetric Gaussian states}\label{secnuma}

Throughout the previous Section we investigated the genuine multipartite entanglement $G_\tau^{res}$ in states of the form $\sig^{(N)}_{K:\{m_1,\ldots,m_K\}}$, as the entanglement shared by $K$ molecules which are sub-blocks of a fully symmetric, pure or mixed Gaussian state. Although the sizes of the molecules were in general different, the severe symmetry constraint on the original state might induce to conclude that the strong monogamy of Gaussian entanglement is a feature limited to permutation-invariant states\footnote{Notice however that a family of non-symmetric four-mode Gaussian states, not related to the classes of states considered here and exhibiting an unlimited entanglement promiscuity, has also been proven to obey the strong monogamy constraint \cite{unlim}.}. Actually, the unitary localizability of multipartite entanglement (see Sec.~\ref{secuniloc}), is not only a key ingredient for the determination of $G_\tau^{res}$ via \eq{monofortemolecolare}, but also enables an immediate re-interpretation of the previous calculations in terms of the genuine multipartite entanglement shared by $K$ single modes (nuclei) in effective non-symmetric pure or mixed $K$-mode Gaussian state $\gr{\tilde{\sigma}}'^{(N)}_{K:\{m_1,\ldots,m_K\}}$, \eq{uniloccm}.

In particular, in our investigations $\sig$ is the CM of fully symmetric Gaussian states, whose symplectic properties are known \cite{adescaling,unitarily}. As a consequence, the nuclear state $\gr{\tilde{\sigma}}'$ shares the distinctive property that all of its symplectic eigenvalues but one are equal to $1$ and the non-unit symplectic eigenvalue is trivially equal to $\sqrt{\det{\gr{\tilde{\sigma}}'}}$, i.e. to the inverse of the purity of the state. States of this form can be regarded as partially saturating the Heisenberg uncertainty relations \ineq{bonfide} and are thus the direct multipartite generalization of the two-mode GLEMS \cite{prl} already introduced in Sec.~\ref{secmol}. We will therefore refer to state $\gr{\tilde{\sigma}}^{(N)}_{K:\{m_1,\ldots,m_K\}}$ as a ``multi-GLEMS'' or $K$-mode GLEMS.

The standard form of multi-GLEMS corresponds to a $K$-mode CM $\gr{\tilde{\sigma}}^{(N)}_{K:\{m_1,\ldots,m_K\}}$ where the reduced CM of any two modes $i$ and $j$ denotes a GLEMS in standard form \cite{prl}, parameterized by the local purity $1/a$ of mode $i$, the local purity $1/b$ of mode $j$, and the global purity $1/c$ of the two-mode reduced state:
\begin{equation}\label{glems}
\sig_{_{\rm GLEMS}}=\left(
\begin{array}{llll}
 a & 0 & g_+ & 0 \\
 0 & a & 0 & g_- \\
 g_+ & 0 & b & 0 \\
 0 & g_- & 0 & b
\end{array}
\right)\,,
\end{equation}
with $g_{\pm} = \big\{\sqrt{[(a-b)^2-(c-1)^2] [(a-b)^2-(c+1)^2]}\pm \sqrt{[(a+b)^2-(c-1)^2] [(a+b)^2-(c+1)^2]}\big\} \big/ \big[4\sqrt{ab}\big]$.
The three purities are in turn related to the determinants of the reduced states of the molecules ${\cal M}_i$ and ${\cal M}_j$ (made of $m_i$ and $m_j$ modes respectively) of the original fully symmetric state  $\sig^{(N)}_{K:\{m_1,\ldots,m_K\}}$, via \eq{abc}. A {\sl Mathematica} instruction to construct the standard form CM $\gr{\tilde{\sigma}}^{(N)}_{K:\{m_1,\ldots,m_K\}}$ of multi-GLEMS --- obtained from unitary localization applied to fully symmetric states $\sig^{(N)}_{K:\{m_1,\ldots,m_K\}}$ --- given $N$, $K$, and a partition $\{m_1,\ldots,m_K\}$, is given by Eq.~(A.10) in the Appendix.
To provide an illustrative example, the following CM describes a mixed $5$-mode multi-GLEMS obtained by starting with a pure fully symmetric $20$-mode Gaussian state, \eq{fscm}, with squeezing $r=1.0$; discarding $5$ modes; partitioning the remaining $15$ modes into molecules, all of different size; and, finally, performing the unitary localization:

\medskip

\begin{widetext}
\[\left.{\gr{\tilde{\sigma}}'^{(20)}_{5:\{1,2,3,4,5\}}}\right|_{r=1}=
\left(
\begin{array}{llllllllll}
 1.871 & 0 & 1.658 & 0 & 1.834 & 0 & 1.956 & 0 & 2.044 & 0 \\
 0 & 1.871 & 0 & -0.1587 & 0 & -0.2152 & 0 & -0.2691 & 0 & -0.3217 \\
 1.658 & 0 & 2.395 & 0 & 2.232 & 0 & 2.380 & 0 & 2.488 & 0 \\
 0 & -0.1587 & 0 & 2.395 & 0 & -0.3535 & 0 & -0.4421 & 0 & -0.5287 \\
 1.834 & 0 & 2.232 & 0 & 2.776 & 0 & 2.633 & 0 & 2.752 & 0 \\
 0 & -0.2152 & 0 & -0.3535 & 0 & 2.776 & 0 & -0.5996 & 0 & -0.7169 \\
 1.956 & 0 & 2.380 & 0 & 2.633 & 0 & 3.069 & 0 & 2.934 & 0 \\
 0 & -0.2691 & 0 & -0.4421 & 0 & -0.5996 & 0 & 3.069 & 0 & -0.8966 \\
 2.044 & 0 & 2.488 & 0 & 2.752 & 0 & 2.934 & 0 & 3.296 & 0 \\
 0 & -0.3217 & 0 & -0.5287 & 0 & -0.7169 & 0 & -0.8966 & 0 & 3.296
\end{array}
\right).\]
\end{widetext}
This state is completely non-symmetric (the reduced single-mode purities are all different) and it is characterized by a global purity $\mu \approx 0.30$ and a genuine multipartite entanglement shared by the $5$ modes equal to $G_\tau^{res} \approx 0.014$, as analytically computable from \eq{monofortemolecolare} [coded as Eq.~(A.9)], given the properties of the original, un-localized state $\sig^{(20)}_{5:\{1,2,3,4,5\}}$.

Thanks to the mechanism of unitary localization, we will now illustrate how the results of the previous Section yield in fact strong evidence that non-symmetric (mixed) Gaussian states in the form of multi-GLEMS obey the strong monogamy decomposition \eq{strongmono} and that $G_\tau^{res}$ is a proper quantifier of genuine multipartite entanglement even for Gaussian states that do not satisfy the constraint of permutation invariance. To substantiate this statements, we review and put in perspective some of the results obtained in the previous Section. The genuine $K$-partite entanglement shared by $K$ modes of a multi-GLEMS increases with the mixedness of each individual mode (these quantities are monotonically increasing functions of the squeezing $r$), similarly to what happens for pure-state bipartite entanglement in general quantum systems; it decreases with the global mixedness of the state; it increases with increasing symmetry among the different modes (i.e. with increasing similarity between each local mixedness). Concerning the scale invariance of the genuine multipartite entanglement, it can be best understood by noting that the fully symmetric $N$-mode state $\sig^{(N)}_{K:\{m_1,\ldots,m_K\}}$ and the re-scaled $(sN)$-mode one $\sig^{(s
N)}_{K:\{s\,m_1,\ldots,s\,m_K\}}$ correspond to the same
effective $K$-mode unitarily localized multi-GLEMS $\gr{\tilde{\sigma}}'^{(N)}_{K:\{m_1,\ldots,m_K\}}$, \eq{uniloccm}, apart from the additional,
irrelevant, single-mode thermal states which are obviously
$s$-degenerate in the latter case.

\begin{figure}[t!]
\includegraphics[width=8.5cm]{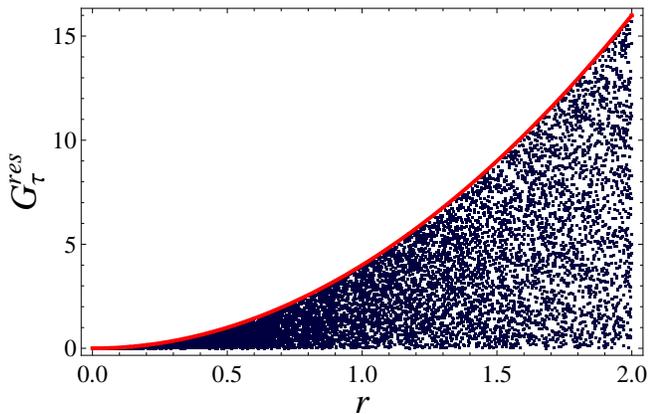} \caption{(Color online)
Plot, as a function of the squeezing $r$, of the genuine
multipartite entanglement $G^{res}_\tau$ in 10~\!000 random fully symmetric $N$-mode Gaussian states. Each point corresponds to the $K$-partite entanglement among $K$ molecules
${\cal M}_1,\ldots,{\cal M}_K$, each comprising $m_j$ modes, with
$\sum_{j=1}^K m_j =M$. For each instance, the real parameter $r$ is
randomized between $0$ and $2$, the integer $N$ (total number of
modes of a pure fully symmetric Gaussian state with squeezing $r$)
is randomized between $2$ and $100$, the integer $M$ (effective
number of modes partitioned in molecules) is randomized between $2$
and $N$, and the integer $K$ (number of molecules) is randomized
between $2$ and $M$. For any $M<N$ the multipartite entanglement is
being computed in a mixed state (as $(N-M)$ modes are traced out);
furthermore, for any $K<M$ the molecules have in general different
sizes. The partition $\{m_1,\ldots,m_K\}$ is chosen at random as
well among all the possible partitions.
Upon application of the unitary localization, each point equivalently corresponds to the genuine
$K$-partite entanglement of generally mixed, generally non-symmetric random $K$-mode Gaussian states belonging to the class of multi-GLEMS. In up to two million states (not
shown here) randomly generated according to this prescription, with $N$ up to $10^3$, all values of $G^{res}_\tau$ have been found non-negative. This numerical evidence supports the conjecture that the strong monogamy decomposition holds for all Gaussian states beyond symmetry constraints under permutation of the modes, and that the positive-defined residual contangle is a proper quantifier of genuine multipartite entanglement for general states on harmonic lattices. The solid line in the figure corresponds to the
instance $M=N=2$, $K=2$, $m_1=m_2=1$, i.e. to the bipartite entanglement in two-mode squeezed states. As the genuine multipartite entanglement decreases with $N$, such a bipartite entanglement stands as an absolute upper limit for any $(N>2)$-mode instance. This bound can be only attained for $N=M=2s$, $K=2$, $m_1=m_2=s$ (with $s$ an integer scale factor) by virtue of the scale invariance discussed in Sec.~\ref{secmolscale}. All the quantities plotted are dimensionless.} \label{listone}
\end{figure}

We conclude our discussion by reporting the results of extended numerical tests on strong monogamy, i.e. on the non-negativity of the genuine multipartite entanglement [\ineq{strongineq}],
as thoroughly illustrated and discussed in Fig.~\ref{listone}. The obtained results can be equivalently regarded in terms of molecular block entanglement in fully symmetric (pure or mixed) Gaussian states, or entanglement shared by single modes in non-symmetric (pure or mixed) multi-GLEMS. Adopting the first picture for simplicity, we  generated  states of the form $\sig^{(N)}_{K:\{m_1,\ldots,m_K\}}$, where all parameters were randomized, as explained in detail in the caption of Fig.~\ref{listone}. The fact that in all instances we found a well defined, non-negative $G_\tau^{res}$, provides good evidence in support of the conjecture that \eq{monofortemolecolare} is a proper quantifier of genuine multipartite entanglement, and that entanglement distributes according to a strong monogamy law of the form \eq{strongmono}, independently of symmetry constraints under permutation of the modes. A full analytic proof of this claim is however, to date, available only in the special case of the multipartite entanglement shared by single modes in states $\sig^{(N)}_{K:\{1,\ldots,1\}}$ \cite{strong}, and it is still lacking in the general case of distributed molecular entanglement.

\section{Concluding remarks}\label{secconcl}

In the present work we have discussed the characterization and quantification of genuine multipartite entanglement in fully inseparable Gaussian states of multimode, bosonic CV systems. We adopted the framework introduced in Ref. \cite{strong} in order to establish some generalizations of the conventional monogamy inequalities on distributed entanglement and to define a proper and computable quantifier of genuine multipartite entanglement. Our main aim has been to verify whether general Gaussian states, not endowed in principle with special symmetries with respect to mode permutation and/or not necessarily partitioned into parties each owning a single mode, obey strong monogamy constraints, i.e. that hold {\em simultaneously} on bipartite and genuine multipartite entanglement. This type of analysis is motivated by conceptual questions as well as by the practical utility of multipartite Gaussian resources for CV quantum information and communication. Such a study is thus of value in order to single out the most efficient ways to engineer and partition Gaussian states yielding optimal performances in quantum information tasks.

In this paper we did not present a comprehensive, rigorous, and conclusive analysis on the structure of multipartite entanglement sharing and on the quantification of genuine multipartite entanglement in general Gaussian states. On the other hand, we obtained a series of partial results that, in our opinion, points to a unified framework and realizes concrete steps towards a positive answer to the questions posed. To summarize, we considered a class of Gaussian states $\sig^{(N)}_{K:\{m_1,\ldots,m_K\}}$, with the following properties. They are $M$-mode fully symmetric mixed Gaussian states (i.e. invariant under permutation of any two modes) obtained from $N$-mode fully symmetric pure Gaussian states by discarding $N-M$ modes. These states are partitioned into $K$ parties, or molecules, respectively formed by $m_1,\ldots,m_K$ modes (with $\sum_{j=1}^K m_j \equiv M$. We derived an analytical formula --- and we included a {\sl Mathematica} code to evaluate it numerically in arbitrary cases--- for the genuine multipartite entanglement (quantified by the residual contangle $G_\tau^{res}$ emerging from the strong monogamy decomposition) shared by the $K$ parties. We characterized various properties of this quantity that we list here synoptically in order to provide a final comprehensive summary:

\begin{itemize}
  \item[(i)] The genuine multipartite entanglement among $K$ ``molecules'' of modes with fixed sizes $m_j$ and assigned partition decreases if the global number of modes $N$ of the original pure state is increased (resulting in a more mixed state of the $M$ modes on which the partition is imposed).
  \item[(ii)] The genuine multipartite entanglement is invariant under global re-scaling of the system, i.e. varying  by the same scale factor both the number of modes $N$ and the individual sizes $m_j$ of each molecule, at fixed number $K$ of molecules.
  \item[(iii)] For a fixed $N$ and fixed number $M$ of non-discarded modes, the genuine multipartite entanglement decreases with increasing number $K$ of molecules in which the $M$ modes are partitioned.
  \item[(iv)] For fixed $N$, $M$, and $K$, the genuine multipartite entanglement exhibits a hierarchical structure with respect to the individual sizes of the molecules $m_j$. It appears to increase with increasing ranking (order index) of the sorted list of all possible partitions of $M$ modes into $K$ molecules. Roughly speaking, it increases as the sizes of individual molecules get closer.
  \item[(v)] The genuine multipartite entanglement always increases with the average squeezing $r$ for fixed values of the other parameters. For any multi-partition of globally pure states ($M=N$) the multipartite entanglement diverges in the limit of infinite squeezing, while for any multi-partition of globally mixed states ($M<N$) it saturates to a finite value. In all states of the considered class there is thus a promiscuous entanglement distribution, i.e. a coexistence between bipartite and any form of multipartite entanglement, with all contributions being  increasing functions of each other.
  \item[(vi)] The genuine multipartite entanglement among $K$ molecules with, generally different, sizes $m_1,\ldots,m_K$ in pure or mixed fully symmetric Gaussian states, is unitarily localizable, i.e. it is equal to the genuine multipartite entanglement among $K$ modes in pure or mixed, generally non-symmetric, Gaussian states with a well defined symplectic spectrum characterized by all symplectic eigenvalues equal to unity but one.
  \item[(vii)] The genuine multipartite entanglement is found to be non-negative in all the instances analyzed, and after extensive numerical tests on ample classes of randomly generated permutation-invariant CMs. This result provides evidence that the strong monogamy constraint holds true both for the bipartite and the genuine multipartite entanglement distributed among blocks of modes of arbitrary size in fully symmetric Gaussian states, and among single modes in the subset of non-symmetric Gaussian states constituted by mixed states of partial minimum uncertainty.
\end{itemize}

Property (vi) stems from a more general result, of independent interest, demonstrated in this paper, namely the unitary localization of bipartite and multipartite correlations in multi-symmetric Gaussian states (i.e. states that are locally invariant under permutation of modes within each molecule, but are in  general not invariant under permutation of two modes taken from different molecules), which extends an analogous result proved previously in the bipartite instance \cite{unitarily}.

The above findings, together with the operational interpretation established in Ref.~\cite{strong} that relates $G_\tau^{res}$ to the optimal fidelity of CV teleportation networks \cite{network,telepoppy} employing fully symmetric Gaussian resources, yield further support to the use of the generalized residual contangle $G_\tau^{res}$ for the quantification of genuine multipartite entanglement in Gaussian states.

Whether the strong monogamy of bipartite and genuine multipartite entanglement is a property that applies
in general to all quantum states remains an open question. From the present results in the Gaussian CV scenario, and in particular from point (vii) above, one is lead to conclude that the constraints on entanglement imposed  by the strong monogamy have a wider domain of validity than de~Finetti-type bounds \cite{definetti,wolfdef} which apply only to fully permutation-invariant states, even though the monogamy-based bounds carry the drawback of being crucially dependent on the actual measure of entanglement that is chosen. An investigation of the strong monogamy decomposition and an evaluation of the corresponding quantifier of genuine multipartite entanglement on states of finite-dimensional systems would provide a relevant test ground in order to verify the general validity of the property of strong monogamy of distributed quantum correlations.

\acknowledgments

We acknowledge financial support from MIUR under PRIN National
Project 2005, INFN, CNR-INFM Coherentia, and ISI Foundation.

\appendix*

\section{Mathematica code}

This appendix lists some of the routines we defined in \textsl{Wolfram
Mathematica 6.0} and used for the evaluation of the genuine
multipartite entanglement shared by modes and/or molecules in fully
symmetric Gaussian states. A brief comment accompanies each
instruction; more details can be retrieved in the main text, in the
specific part in which the corresponding instruction is referred to.
The following two packages need to be loaded in the {\sl Mathematica}
kernel:

\noindent\(\\ \pmb{<<\text{LinearAlgebra$\grave{ }$MatrixManipulation$\grave{ }$}}\\
\pmb{<<\text{Combinatorica$\grave{ }$}}\\ \)

\begin{itemize}\item[(A.1)]
Definition of contangle in terms of the squared determinant $
d^2$ involved
in the computation of the Gaussian entanglement measures [see \eq{tau}]:
\end{itemize}
\noindent\(\pmb{\text{contangle}[\text{\textit{d2}}\_] :=
(\text{ArcSinh}[\text{Sqrt}[-1 + \text{\textit{d2}}]])^2;}\\\)

\begin{itemize}
\item[(A.2)] Covariance matrix of fully symmetric pure Gaussian states of
$n$ modes:
   \end{itemize}
\noindent\(\pmb{\text{FulSymCM}[n\_]:=\text{BlockMatrix}[\text{Table}[}\\
\pmb{\text{If}\bigg[i==j,\left(
\begin{array}{cc}
 b & 0 \\
 0 & b
\end{array}
\right),\bigg.} \\
\pmb{\left(1/(2b(n-1))\left(
\begin{array}{cc}
 (\text{\textit{k1}}+\text{\textit{k2}}) & 0 \\
 0 & (\text{\textit{k1}}-\text{\textit{k2}})
\end{array}
\right)\text{/.}\right.}\\
\pmb{\left\{\text{\textit{k1}}\to 2+b^2 (-2+n)-n,\text{\textit{k2}}\to \right.}\\
\pmb{\bigg.\left.\left.\text{Sqrt}\left[\left(-1+b^2\right)\left(b^2 n^2-(n-2)^2\right)\right]\right\}\right)\bigg],}\\
\pmb{\{i,1,n\},\{j,1,n\}]];}\\\)

\begin{itemize}
     \item[(A.3)]
     Relation between single-mode determinant $b^2$ and squeezing $r$ of pure, fully symmetric $n$-mode Gaussian states, as a function of
     $n$:
        \end{itemize}
 \noindent\(\pmb{\text{mybsq}[n\_]:=\frac{1}{n^2}(2+(-2+n) n+2
(-1+n) \text{Cosh}[4 r]);}\\\)

\begin{itemize}
     \item[(A.4)]
Reduced determinant of a block of $l$ modes out of an $n$-mode fully symmetric pure Gaussian state:
\end{itemize}
\noindent\(\pmb{\text{ghredet}[n\_,l\_]:=}\\
\pmb{(1/(n-1))*\left(l(n-l)b^2-(l-1)(n-l-1)\right);}\)
\\ \\
\noindent\(\pmb{\text{ghd}[n\_,l\_]:=}\\
\pmb{\text{Refine}[\text{Simplify}[\text{ghredet}[n,l]\text{/.}b\to \text{Sqrt}[\text{mybsq}[n]]],}\\
\pmb{r>0]\text{//}\text{Simplify};}\\\)

\begin{itemize}\item[(A.5)]
Gaussian entanglement measure (squared determinant $d^2$) for a two-mode
GLEMS with local purities $1/a$ and $1/b$ and global purity $1/c$:
\end{itemize}

\noindent\(\pmb{\text{d2glems}[a\_,b\_,c\_]:=}\\
\pmb{\text{If}\left[ c\geq \sqrt{-1+a^2+b^2},1,\text{If}[c<\text{Sqrt}[\right.}\\
\pmb{\frac{1}{2 \left(a^2+b^2\right)}\left(2 \left(a^2+b^2\right)+\left(a^2-b^2\right)^2 +\text{Abs}\left[a^2-b^2\right] \right.}\\
\pmb{\left.\left.\surd \left(\left(a^2-b^2\right)^2+8\left(a^2+b^2\right)\right)\right)\right] ,\frac{\left(a^2-b^2\right)^2}{\left(-1+c^2\right)^2},}\\
\pmb{\frac{1}{8 c^2}\left(-a^4-b^4-\surd ((-1+a-b-c) (1+a-b-c) \right.}\\
\pmb{(-1+a+b-c) (1+a+b-c)) \surd ((-1+a-b+c) }\\
\pmb{(1+a-b+c) (-1+a+b+c) (1+a+b+c))-}\\
\pmb{\left.\left.\left.\left(-1+c^2\right)^2+2 b^2
\left(1+c^2\right)+2 a^2
\left(1+b^2+c^2\right)\right)\right]\right];}\\\)

\begin{itemize}\item[(A.6)]
Gaussian entanglement measure (squared determinant $d^2$) for a $(i\vert
j)$-mode bipartition in a $n$-mode pure fully symmetric Gaussian
state:
\end{itemize}
\noindent\(\pmb{\text{d2ijn}[n\_,i\_,j\_]:=}\\
\pmb{\text{Refine}[(((\text{Refine}[(\text{d2glems}[\text{Sqrt}[\text{ghredet}[n,i]],}\\
\pmb{\text{Sqrt}[\text{ghredet}[n,j]],\text{Sqrt}[}\\
\pmb{\text{ghredet}[n,i+j]]]\text{//}\text{Simplify}),b>1])\text{/.}}\\
\pmb{b\to \text{Sqrt}[\text{mybsq}[n]])\text{//}\text{Simplify}),r>0];}\\
\)

\begin{itemize}\item[(A.7)]
Genuine $k$-partite entanglement shared by $k$ individual modes in a
$n$-mode pure fully symmetric state, as a function of $k$, $n$,
and the squeezing $r$ (as computed in \cite{strong}):
\end{itemize}
\noindent\(\pmb{\text{GEnt}[k\_,n\_,r\_]:=}\\
\pmb{\text{Sum}\left[(-1)^j \text{ArcSinh}\left[\left.\left(2 \sqrt{-1-j+k} \text{Sinh}[2 r]\right)\right/\right.\right.}\\
\pmb{\left.\left(\sqrt{n} \sqrt{-j+e^{4 r} (j+n-k)+k}\right)\right]^2 }\\
\pmb{\text{Binomial}[-1+k,j],\{j,0,k-2\}];}\\\)

\begin{itemize}\item[(A.8)]
Sorted list of all the possible partitions of $n$ modes into $k$
molecules:
\end{itemize}
\noindent\(\pmb{\text{KSortedPartitions}[n\_,k\_]:=\text{Module}[}\\
\pmb{\{\text{\textit{tabpart}},\text{\textit{tabpartsort}},u\},\text{\textit{tabpart}}=\text{Partitions}[n];}\\
\pmb{\text{\textit{tabpartsort}}=\{\};\text{Do}[\text{If}[\text{Length}[\text{\textit{tabpart}}[[u]]]==k,}\\
\pmb{\text{AppendTo}[\text{\textit{tabpartsort}}, \{\{\text{Sort}[\text{\textit{tabpart}}[[u]]]\}\}]],}\\
\pmb{\{u,\text{Length}[\text{\textit{tabpart}}]\}];}\\
\pmb{\text{Return}[\text{MatrixForm}[\text{Sort}[\text{\textit{tabpartsort}}]]]];}\\\)

\begin{itemize}\item[(A.9)]
Genuine multipartite entanglement shared by $k$ molecules made of
\textit{mpart} = $\{$\(m_1\), $\ldots $, \(m_k\)$\}$ modes
respectively \( \left(\sum
_{\text{\textit{$j$}}}\text{\textit{$m_j$}}\leq
\text{\textit{$n$}}\right)\), in a global $n$-mode pure fully symmetric Gaussian state (recursive definition):
\end{itemize}

\noindent\(\pmb{\text{GMolEnt}[n\_,k\_,\text{\textit{mpart}}\_]:=}\\
\pmb{\text{contangle}[\text{d2ijn}[n,\text{\textit{mpart}}[[1]],}\\
\pmb{\text{Sum}[\text{\textit{mpart}}[[i]],\{i,2,k\}]]]-\text{Sum}[}\\
\pmb{\text{Sum}[\text{GMolEnt}[n,m,\text{KSubsets}[\text{\textit{mpart}},m][[l]]],}\\
\pmb{\{l,\text{Binomial}[k-1,m-1]\}],\{m,k-1,2,-1\}];}\\\)

\begin{itemize}\item[(A.10)]
Covariance matrix of the equivalent $k$-mode state (multi-GLEMS)
associated, via unitary localization, to a multi-partition of $n$
modes into $k$ molecules made of \textit{mpart} = $\{$\(m_1\),
$\ldots $, \(m_k\)$\}$ modes respectively \( \left(\sum
_{\text{\textit{$j$}}}\text{\textit{$m_j$}}\leq
\text{\textit{$n$}}\right)\):
\end{itemize}

\noindent\(\pmb{\text{UniLocCM}[n\_,k\_,\text{\textit{mpart}}\_]:=}\\
\pmb{\text{Refine}\left[\left(\text{Refine}\left[\left(\text{BlockMatrix}\left[\left(\text{Table}\left[\text{If}\left[i==j,\alpha _i,\right.\right.\right.\right.\right.\right.\right.\right.}\\
\pmb{\left.\left.\left.\left.\text{If}\left[i<j,\epsilon _{i,j},\epsilon _{j,i}\right]\right],\{i,k\},\{j,k\}\right]\right)\right]\text{/.}}\\
\pmb{\text{Join}\left[\text{Table}\left[\alpha _i\to \text{Sqrt}[\text{ghd}[n,\text{\textit{mpart}}[[i]]]]\right.\right.}\\
\pmb{\text{IdentityMatrix}[2],\{i,k\}],}\\
\pmb{\text{Flatten}\left[\text{Table}\left[\epsilon _{i,j}\to \text{DiagonalMatrix}[\right.\right.}\\
\pmb{\left\{\frac{1}{4 \sqrt{a_i a_j}}\left(\surd \left(\left(-\left(-1+a_{\text{\textit{ipj}}}\right){}^2+\left(a_i-a_j\right){}^2\right) \right.\right.\right.}\\
\pmb{\left.\left(-\left(1+a_{\text{\textit{ipj}}}\right){}^2+\left(a_i-a_j\right){}^2\right)\right)+}\\
\pmb{\surd \left(\left(-\left(-1+a_{\text{\textit{ipj}}}\right){}^2+\left(a_i+a_j\right){}^2\right) \right.}\\
\pmb{\left.\left.\left(-\left(1+a_{\text{\textit{ipj}}}\right){}^2+\left(a_i+a_j\right){}^2\right)\right)\right),}\\
\pmb{\frac{1}{4 \sqrt{a_i a_j}}\left(\surd \left(\left(-\left(-1+a_{\text{\textit{ipj}}}\right){}^2+\left(a_i-a_j\right){}^2\right) \right.\right.}\\
\pmb{\left.\left(-\left(1+a_{\text{\textit{ipj}}}\right){}^2+\left(a_i-a_j\right){}^2\right)\right)-}\\
\pmb{\surd \left(\left(-\left(-1+a_{\text{\textit{ipj}}}\right){}^2+\left(a_i+a_j\right){}^2\right) \right.}\\
\pmb{\left.\left.\left.\left(-\left(1+a_{\text{\textit{ipj}}}\right){}^2+\left(a_i+a_j\right){}^2\right)\right)\right)\right\}\text{/.}}\\
\pmb{\left\{a_i\to \text{Sqrt}[\text{ghd}[n,\text{\textit{mpart}}[[i]]]],a_j\to \right.}\\
\pmb{\text{Sqrt}[\text{ghd}[n,\text{\textit{mpart}}[[j]]]],a_{\text{\textit{ipj}}}\to \text{Sqrt}[}\\
\pmb{\text{ghd}[n,\text{\textit{mpart}}[[i]]+\text{\textit{mpart}}[[j]]]]\}],}\\
\pmb{\{i,k\},\{j,k\}]]]),r>0]\text{//}}\\
\pmb{\text{Simplify}),r>0]\text{//}\text{Simplify};}\\\)



\end{document}